\documentclass[a4paper,fleqn]{cas-sc}

\makeatletter
\renewcommand{\section}{\@startsection{section}{1}{\z@}
	{24\p@ \@plus 6\p@ \@minus 2\p@}
	{6\p@}
	{\normalfont\normalfont\normalsize\bfseries}} 

\renewcommand{\subsection}{\@startsection{subsection}{2}{\z@}
	{18\p@ \@plus 4\p@ \@minus 2\p@}
	{4\p@}
	{\normalfont\normalsize\itshape}} 
\makeatother

\usepackage{graphicx}
\usepackage{amsmath}
\usepackage{amssymb}
\usepackage{booktabs}
\usepackage[sort, numbers,authoryear,longnamesfirst]{natbib}
\setcitestyle{numbers,square}
\usepackage{amssymb}
\usepackage[utf8]{inputenc} 
\usepackage[T1]{fontenc}    
\usepackage{hyperref}       
\usepackage{url}            
\usepackage{booktabs}       
\usepackage{amsfonts}       
\usepackage{nicefrac}       
\usepackage{microtype}      
\usepackage{xcolor}
\usepackage{tikz}
\usepackage{tabularx}
\usepackage{makecell}
\usepackage{utfsym}
\usepackage{hyperref}
\usepackage[font=rm]{caption}

\usepackage{bm}
\usepackage{url}            
\usepackage{nicefrac}       
\usepackage{microtype}      
\usepackage{tabu}
\usepackage{multirow}       
\usepackage{graphicx}
\usepackage{graphics}
\usepackage{tabularx}
\usepackage{makecell}
\usepackage{caption}
\usepackage{wrapfig}
\usepackage{enumitem}
\usepackage{subcaption}
\usepackage{algorithm}
\usepackage{arydshln}
\usepackage{algpseudocode}
\usepackage{color, colortbl}
\usepackage{listings}
\definecolor{Gray}{gray}{0.95}
\definecolor{orange}{rgb}{0.9,0.5,0}

\usepackage{pifont}
\usepackage[capitalize]{cleveref}
\crefname{section}{Sec.}{Secs.}
\Crefname{section}{Section}{Sections}
\Crefname{table}{Table}{Tables}
\crefname{table}{Tab.}{Tabs.}
\usepackage[authoryear,longnamesfirst]{natbib}

\def\tsc#1{\csdef{#1}{\textsc{\lowercase{#1}}\xspace}}
\tsc{WGM}
\tsc{QE}

\begin{document}
\let\WriteBookmarks\relax
\def\floatpagepagefraction{1}
\def\textpagefraction{.001}

\shorttitle{}    

\shortauthors{}  

\title [mode = title]{Efficient Feedback Gate Network for Hyperspectral Image Super-Resolution}

%

\author[1]{Xufei Wang}[orcid=0009-0006-8442-9526]
\ead{wangxf024@gmail.com}
\credit{}
\affiliation[1]{organization={School of Electronic and Information Engineering, Anhui University},
            addressline={No 111, Jiulong Road}, 
            city={Hefei},
            postcode={230039}, 
            state={Anhui},
            country={China}}

\author[1]{Mingjian Zhang}[orcid=0000-0002-7731-6113]
\ead{zhang9317112@gmail.com}
\credit{}

\author[1]{Fei Ge}[orcid=0009-0009-1187-9691]
\ead{gefei0921@outlook.com}
\credit{}

\author[1]{Jinchen Zhu}[]
\ead{jinchen.z@outlook.com}
\credit{}

\author[2]{Wen Sha}[]
\ead{04069.ahu.edu.cn}
\credit{}
\affiliation[2]{organization={School of Artificial intelligence, Anhui University},
	addressline={No 111, Jiulong Road}, 
	city={Hefei},
	postcode={230039}, 
	state={Anhui},
	country={China}}

\author[1]{Jifen Ren}[]
\ead{P42314014@stu.ahu.edu.cn}
\credit{}

\author[1]{Zhimeng Hou}[]
\ead{3171376528@qq.com}
\credit{}

\author[3]{Shouguo Zheng}[]
\ead{zhshg1985@163.com}
\credit{}
\affiliation[2]{organization={Institutes of Physical Science, Chinese Academy Sciences},
	addressline={}, 
	city={Hefei},
	postcode={230031}, 
	state={Anhui},
	country={China}}

\author[1]{ling Zheng}[]
\cormark[1]
\ead{zhengling@ahu.edu.cn}
\credit{}

\author[1]{Shizhuang Weng}[orcid=0000-0002-7147-8496]
\cormark[1]
\ead{weng_1989@126.com}
\credit{}

\cortext[1]{Corresponding author at: School of Electronic and Information Engineering, Anhui University, Hefei, 230039, Anhui, China.}


\begin{abstract}
Even without auxiliary images, single hyperspectral image super-resolution (SHSR) methods can be designed to improve the spatial resolution of hyperspectral images. However, failing to explore coherence thoroughly along bands and spatial–spectral information leads to the limited performance of the SHSR. In this study, we propose a novel group-based SHSR method termed the efficient feedback gate network, which uses various feedbacks and gate operations involving large kernel convolutions and spectral interactions. In particular, by providing different guidance for neighboring groups, we can learn rich band information and hierarchical hyperspectral spatial information using channel shuffling and dilatation convolution in shuffled and progressive dilated fusion module(SPDFM). Moreover, we develop a wide-bound perception gate block and a spectrum enhancement gate block to construct the spatial–spectral reinforcement gate module (SSRGM) and obtain highly representative spatial–spectral features efficiently. Additionally, we apply a three-dimensional SSRGM to enhance holistic information and coherence for hyperspectral data. The experimental results on three hyperspectral datasets demonstrate the superior performance of the proposed network over the state-of-the-art methods in terms of spectral fidelity and spatial content reconstruction. The code is available at \href{https://github.com/X-F-Wang/EFGN.git.}{https://github.com/X-F-Wang/EFGN.git.}
\end{abstract}


\begin{highlights}
\item Efficient feedback gate network gains excellent SHSR. 
\item Channel shuffle and dilation convolution provide rich channel and spatial information. 
\item Large kernel strip convolution explore image anisotropy and extract global information
\item 3D gate operation refines holistic features of HSI. 
\end{highlights}


\maketitle
\section{Introduction}

In contrast to red, green, and blue three-band images or multispectral images, hyperspectral images can record information covering dozens to hundreds of continuous spectral bands, reflecting the intrinsic and subtle properties of objects \cite{liu2024hyperspectral}. Owing to their excellent capacity, hyperspectral images are widely used in many areas, such as the environment \cite{xu2022hyperspectral}, agriculture, and military. For an acceptable signal-to-noise ratio to be maintained, a tradeoff between the spatial and spectral resolutions of hyperspectral data is inevitable in hardware imaging devices. With respect to the inherent attribute of rich spectral information, hyperspectral images often possess relatively low spatial resolution, which limits their further application.
Super-resolution (SR) is a postprocessing software technique that can infer high-resolution (HR) images from low-resolution (LR) images \cite{he2023single}. This technique is practical for overcoming hardware bottlenecks. Depending on whether auxiliary information, such as panchromatic, RGB or multispectral images, is used, the SRs of hyperspectral images can be classified into two categories: fusion-based hyperspectral image SR and single hyperspectral image super-resolution (SHSR). Although the former achieves considerable performance, coregistered LR images and HR auxiliary images are difficult to obtain in real applications. Nonetheless, HR auxiliary images can mine only the information of LR images to reconstruct these HR images.
The majority of early SHSR methods rely on handcrafted priors, such as self-similarity, sparseness, or low rank \cite{wang2017hyperspectral}. In general, the design of priors is elaborate, labor intensive, and extremely coarse to describe the inherent spatial–spectral properties in hyperspectral images, leading to SR distortion. Recently, the deep convolution neural network (DCNN) was introduced into the SHSR because of its powerful representational ability \cite{chen2024single}. Some emerging traditional DCNN methods in natural image SR, such as the SRCNN \cite{dong2015image} and EDSR \cite{lim2017enhanced}, have been directly applied in the SHSR. These methods mainly adopt two-dimensional (2D) convolution operations to focus on image spatial information, and the spectral correlation of hyperspectral images is difficult to maintain. For the problem of spectral disorder to be alleviated, the two-step methods, such as UCNN \cite{lu2021hyperspectral}, first reconstruct the spatial resolution of the hyperspectral image via a DCNN and use a manual process to further integrate the spectral information. The performance of these methods is highly dependent on the number of iterations and the number of manual process endmembers, which is inflexible and time-consuming. 3D-FCNN subsequently utilized three-dimensional (3D) convolution and its variations to reconstruct hyperspectral images because these convolutions can simultaneously extract spatial and spectral features \cite{mei2017hyperspectral}. Interactformer has introduced transformer structures to obtain powerful global representations to achieve better SHSR performance \cite{liu2022interactformer}. However, 3D convolution and transformers often result in an abundant module burden.
Regarded as a new set of solutions, group-based methods, such as SSPSR \cite{jiang2020learning}, RFSR \cite{9380508}, and CEGATSR \cite{liu2022cnn}, divide hyperspectral images along the spectral dimension into subgroups to execute the respective SRs. As only a few channels exist in the image of each subgroup, the calculation burden and parameters decrease considerably, the SR efficiency is higher than that of the previous methods, and the spectral correlation can be preserved to some extent. However, group-based methods still face some challenges. First, most of the group-based methods neglect the associations of adjacent groups and loosen the coherence of spectral information \cite{wang2020hyperspectral}, which leads to spectral disorder in the reconstructed images. Second, in most existing group-based methods, the convolutions of small kernels and transformers lack efficient feature learning ability to obtain rich and adequate information \cite{wang2023group}.
For the aforementioned problems to be solved, we propose an efficient feedback gate network (EFGN) for group-based SHSR to learn complementary and abundant spatial–spectral relevance. 
Specifically, we introduce Shuffled and progressive dilated fusion module (SPDFM) to enhance the flow of information between branches. In SPDF, channel shuffle can provide rich spectral channel information through parameter-free structural adjustment of channel arrangement, while the dilation convolution provides different levels of spatial information from local to global by changing the dilation rate, and the channel information and spatial information converge in the neighboring groups to provide complementary and diversified a priori guidance to enhance the SR performance.
Then, for adequate representative features to be obtained, we design the spatial–spectral reinforcement gate module (SSRGM). In the SSRGM, we exploit parallel decomposed large-kernel convolution in the form of element multiplication, namely, the gate operation, to gain an enlarged receptive field and feature content, and the spectral information is gathered via the gate operation of channel mixing and channel attention. The 3D-SSRGM was proposed to mine holistic features further. When 3D convolution is used, the potential spatial‒spectral relationships between hyperspectral bands can be explored simultaneously.
Qualitative and quantitative experiments on various hyperspectral datasets demonstrate the superiority of our method over other state-of-the-art methods. In summary, the main contributions of our work are as follows.
\begin{itemize}
	\item[$\bullet$]We propose a group-based SHSR network, the EFGN, which consists mainly of an SPDFM and an SSRGM. The SPDFM is designed to provide various guidance to subgroup image SRs, and the SSRGM gains adequate spatial information and reinforces spectral features.
	\item[$\bullet$]In SPDFM, the channel and spatial prior not only helps to reconstruct the contour and texture features of neighboring subgroups, but also maintains the consistency of the spectral bands.
	\item[$\bullet$]The SSRGM is composed of a wide-bound perception gate block (WPGB) and a spectrum enhancement gate block (SEGB). WPGB utilizes gate integration of large-kernel strip convolutions to exploit spatial information efficiently with a large receptive field; then, SEGB exploits channel gate interactions to increase band consistency. The 3D-SSRGM is applied after subgroup merging to refine the coalescing feature.
\end{itemize}
The remaining sections of this paper are organized as follows: Section II briefly reviews the recent SR methods for hyperspectral images. Section III describes the details of the proposed method. Section IV presents the setting and experiment results. Section V summarizes this study.

\section{Related work}

SHSR has been widely studied in recent years. To better understand our work, we briefly introduce the development of SHSR and review mainstream methods. The content is presented in two parts: deep learning-based SHSR and group-based SHSR method.

\subsection{Deep learning-based SHSR}
Existing deep learning based SR methods for natural images, such as  EDSR\cite{lim2017enhanced}, BSRN \cite{li2022blueprint}, and DiVANet \cite{behjati2023single}, are directly introduced into the SHSR. The methods focus only on improving the spatial resolution, and the spectral consistency is preserved inadequately. This situation easily leads to disastrous performance in the spectral dimension, which ultimately affects the overall SR effect. Such a scenario inevitably induces severe degradation in spectral fidelity, thereby compromising the global super-resolution performance.
Postprocessed SHSRs are initially proposed to address the issues pertaining to large spectral bands of hyperspectral images. SDCNN \cite{li2017hyperspectral} uses 2D convolution to extract spatial features and uses a spatial constraint strategy to preserve spectral consistency. Yuan et al. \cite{yuan2017hyperspectral} implements nonnegative matrix factorization to reconstruct spectral information. However, postprocessing methods inevitably induce errors, are time-consuming and unstable. The methods based on 3D convolution subsequently exploit the synchronous extraction of spatial context and spectral information. Pioneering work is conducted by the 3DFCNN \cite{sanchez20223dfcnn}, which stacks the full sequential 3D convolutions and preserves the spectral correlation of the reconstructed images well. However, the model computational complexity increases explosively. MCNet \cite{li2020mixed} alternately utilizes 2D and 3D convolutions to balance learning ability and complexity. A parallel network that uses a transformer and 3D convolution is subsequently designed as an Interactformer to extract global and local features. SRDNet \cite{10445371} separate 3D convolution to maintain acceptable model burden and design the self-attentive pyramid structure to capture similarity in the spatial domain. The SR results are further improved, but the computational complexity and memory occupation remain high. In summary, these end-to-end methods holistically process large hyperspectral cubes, which leads to a large model burden and inefficient feature extraction.

\subsection{Group-based SHSR method}
GDRRN \cite{li2018single} pioneeringly employs hierarchically nested recursive modules with group-wise convolutional operations to enable sequential channel-wise feature extraction. On this basis, SSPSR \cite{jiang2020learning}, a group-based method, innovatively proposes a framework to group hyperspectral images into subgroups along bands and utilizes a spatial–spectral feature extraction module for the respective SRs. CEGATSR \cite{liu2022cnn} adopts a graph attention block and depthwise separable convolution into group-based methods to obtain local and nonlocal features, and the spatial-channel attention block reweights the importance of different channels to capture valuable information. RFSR \cite{9380508} proposes a recurrent interaction mechanism to supply complementary information with adjacent groups and applies a regularization module to enhance spatial and spectral correlations. SGARDN \cite{liu2021spectral} incorporates a residual dense module and a spectral attention mechanism to extract spatial information effectively and alleviate spectral disorder. Dilated convolution and transformer structures were combined in MSDformer \cite{10252045} to achieve long-range dependence \cite{chen2023msdformer}.
Nevertheless, the current group-based methods face two dilemmas. First, the lack of various information interactions between adjacent groups worsens the spectral coherence in SR images. Second, the feature extraction module suffers from inefficient feature extraction of spatial–spectral information \cite{wang2023hyperspectral}.

\section{Proposed method}

\begin{figure}[htbp]
	\centering
	\begin{subfigure}{1\linewidth}
		\centering
		\includegraphics[width=1\linewidth]{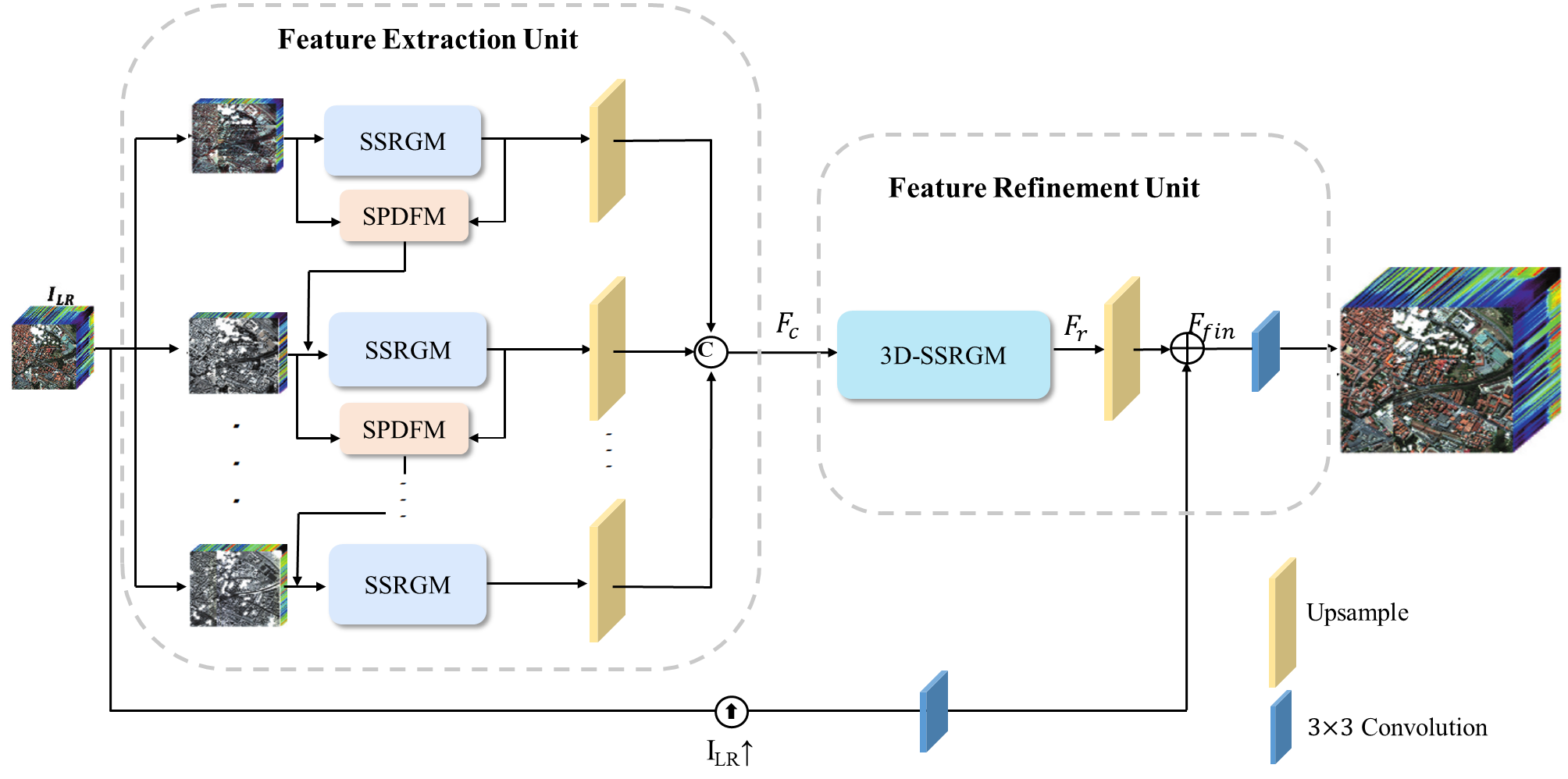}
	\end{subfigure}
	\caption{Overall structure of EFGN.}
	\label{structure}
\end{figure}

In this section, we elaborate on the proposed SHSR method EFGN. The following content is divided into four sections: overall framework, SPDF, SSRGM, and loss function.
\subsection{Overall framework}
Our proposed EFGN implements a group-based framework. Fig. \ref{structure} shows the EFGN, which consists of a feature extraction unit and a feature refinement unit. We denote the LR image as $I_{LR}\in\mathbb{R}^{H \times W \times C}$, the HR image as $I_{HR}\in\mathbb{R}^{sH \times sW \times C}$, and output SR image as $I_{SR}\in\mathbb{R}^{sH \times sW \times C}$, where $H$, $W$, $C$ and $s$ denote the height, width, and channels of the hyperspectral image, and the scale factors of SR. Our goal is to infer $I_{SR}$ from $I_{LR}$ via the EFGN:
\begin{equation}
I_{SR} = H_{EFGN}(I_{LR})
\end{equation}
Where $H_{EFGN}(\cdot)$ denotes the function of EFGN.

Following the mode of group-based methods, we split the input LR image into $G$ groups in a nonoverlapping way, $I_{LR} = [I_{LR}^1, ..., I_{LR}^g, ..., I_{LR}^G]$. Complementary SR guidance must be provided, and the continuity of hyperspectral data must be conserved. Accordingly, our SPDFM aims to bring hierarchical spatial and spectral channel information from the previous group to the current group input. An updated $g_{th}$ group input can be formulated as:
\begin{equation}
\widehat{I_{LR}^g} = H_{SPDFM}(I_{LR}^g, I_{LR}^g-1,F_{e}^g-1)
\end{equation}
Where $\widehat{I_{LR}^g}$, $H_{SPDFM}(\cdot),$ and $F_{e}^g-1$ represent the updated input of the $g_{th}$ group, the SPDFM function, and the extracted enhanced feature of the $g_{th}-1$ group.

The SSRGM is designed to extract the enhanced feature $F_{e}^{g-1}\in\mathbb{R}^{H \times W \times \frac{C}{G}}$ from each updated input $\widehat{I_{LR}^g} \in\mathbb{R}^{H \times W \times \frac{C}{G}}$ and promote the spatial resolution to obtain $F_{p}^{g} \in\mathbb{R}^{\frac{sH}{2} \times \frac{sW}{2} \times \frac{C}{G}}$ at scale factors $\frac{s}{2}$ by executing upsampling, which can be formulated as:
\begin{equation}
\begin{split}
F_{e}^{g} &= H_{SSRGM}(\widehat{I_{LR}^g}),  \\
F_{p}^{g} &= H_{UP}(F_{e}^{g})
\end{split}
\end{equation}
Where $H_{SSRGM}$ denotes the function of the SSRGM, and $H_{UP}$ denotes the upsampling function. We concatenate group feature to form the concentrated feature $F_{c}\in\mathbb{R}^{\frac{sH}{2} \times \frac{sW}{2} \times C}$, which can be formulated as:
\begin{equation}
F_{c} = Cat(F_{p}^1, ..., F_{p}^{g}, ..., F_{p}^{G})
\end{equation}
Where $Cat(\cdot)$ denote the concatenating operation.

We refine the feature $F_{c}$  via 3D-SSRGM to consolidate the band coherence and learn holistic information. This feature can be presented as 
\begin{equation}
F_{r} = H_{3D-SSRGM}(F_{c})
\end{equation}
where $F_{r} \in\mathbb{R}^{\frac{sH}{2} \times \frac{sW}{w} \times C}$ denotes the refined feature, and $H_{3D-SSRGM}$ denotes the function of the 3D-SSRGM.

The feature spatial size is recovered to the HR image size by upsampling. We can obtain the upsampled feature $F_{ur}\in\mathbb{R}^{sH \times sW \times C}$. The process can be formulated as:
\begin{equation}
F_{ur} = H_{up}(F_{r})
\end{equation}

The feature of the bicubic feature is directly added to the network tail to better match the HR feature, which can be formulated as:
\begin{equation}
\begin{split}
&F_{fin} =F_{ur} + H_{c}(I_{LR}\uparrow),  \\
&I_{sr}  = H_{c}(F_{fin})
\end{split}
\end{equation}
Where $F_{fin}\in\mathbb{R}^{sH \times sW \times C}$ represents the final feature, $H_{c}(\cdot)$ denotes the $3 \times 3$ convolution, and the $I_{LR} \uparrow$ s the bicubic upsampling version of $I_{LR}$.

\subsection{Shuffled and progressive dilated fusion module}
\begin{figure}[h!]
	\centering
	\begin{subfigure}{1\linewidth}
		\centering
		\includegraphics[width=0.5\linewidth]{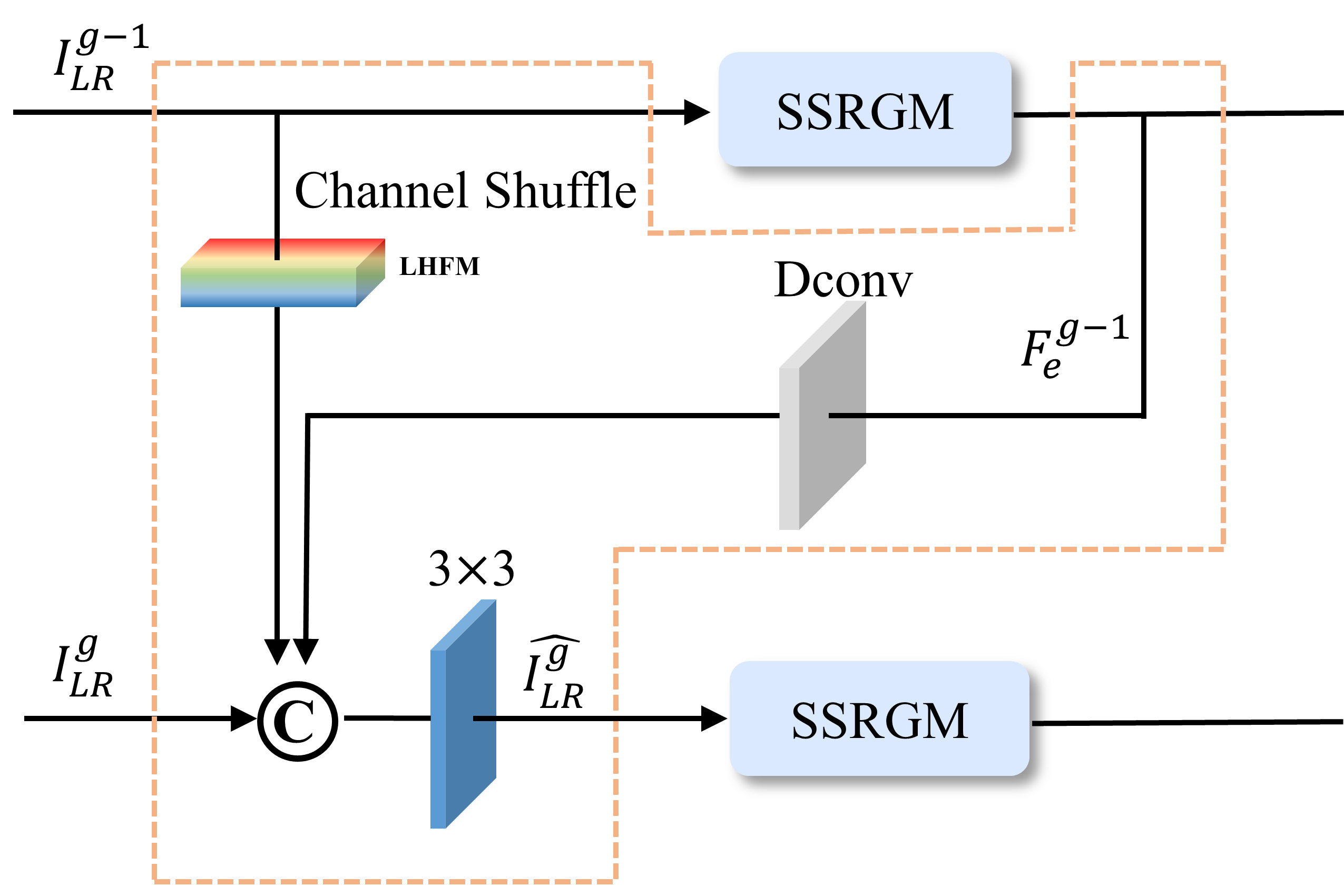}
	\end{subfigure}
	\caption{Structure of the proposed SPDFM.}
	\label{LHFM}
\end{figure}

In order to preserve the spectral continuity, the previous feedback modules, such as RFSR \cite{9380508}, enhance bands relevance by merging unitary information or simply fixing feature in adjacent groups. Wang et al. \cite{wang2023remote} constructed a network structure with different branches to extract information at different levels to improve the diversity of information, thus significantly enhancing the effect of SR. We decide the feedback mechanism between branches to break the independence between branches so that each branch can learn the features extracted by other branches. Additional channels and spatial information are also used to increase the richness of the information learned by all branches.
The structure of SPDF is shown in Fig. \ref{LHFM}. 

For the $g_{th}$ Group, the previous group input $I_{LR}^{g-1}$ contains adjacent original information and is available for primary supplements. Different from spatial shuffle \cite{zhao2024ssir}, we use channel shuffle to divide original feature along channel into pieces, then shuffle these pieces into new feature, which allows the ample information interchange and mapping without extra parameters. We process $I_{LR}^{g-1}$ with channel shuffle to increase the diversity of learnable information and form the spectral channel guidance $F_{C}^{g}\in\mathbb{R}^{H \times W \times \frac{C}{G}}$. The feature processed by the SSRGM $F_{e}^{g-1}\in\mathbb{R}^{H \times W \times \frac{C}{G}}$ is more abstract, sharp, and expressive, which is beneficial for guiding contour reconstruction. Here, we acquire the hierarchical spatial guidance $F_{D}^{g} \in\mathbb{R}^{H \times W \times \frac{C}{G}}$ by handling $F_{e}^{g-1}$ through dilation convolution. We progressively increase dilation rates to capture hierarchical spatial features from local to global contexts. The abovementioned steps can be formulated as:
\begin{equation}
\begin{split}
F_{C}^{g} &= H_{CS}(I_{LR}^{g-1}), \\
F_{D}^{g} &= H_{D}(F_{e}^{g-1}) 
\end{split}
\end{equation}
Where $H_{CS}$ and $H_{D}$ represent the functions of channel shuffle and dilation convolution. For the first group, all feedback information is initialized by zero values of the same size.

Then, we fuse the feedback guidance with the present group LR input $I_{LR}^{g}\in\mathbb{R}^{H \times W \times \frac{C}{G}}$, and the $3 \times 3$ convolution is applied to adjust the number of spectral bands while aligning guidance and features to form the updated input of the $g_{th}$ group $\widehat{I_{LR}^g}\in\mathbb{R}^{H \times W \times \frac{C}{G}}$, which can be formulated as:
\begin{equation}
\widehat{I_{LR}^g} = H_{C}(Cat(I_{LR}^{g},F_{C}^{g},F_{D}^{g})
\end{equation}

\subsection{Spatial–spectral reinforcement gate module}
\begin{figure}[h!]
	\centering
	\begin{subfigure}{1\linewidth}
		\centering
		\includegraphics[width=0.7\linewidth]{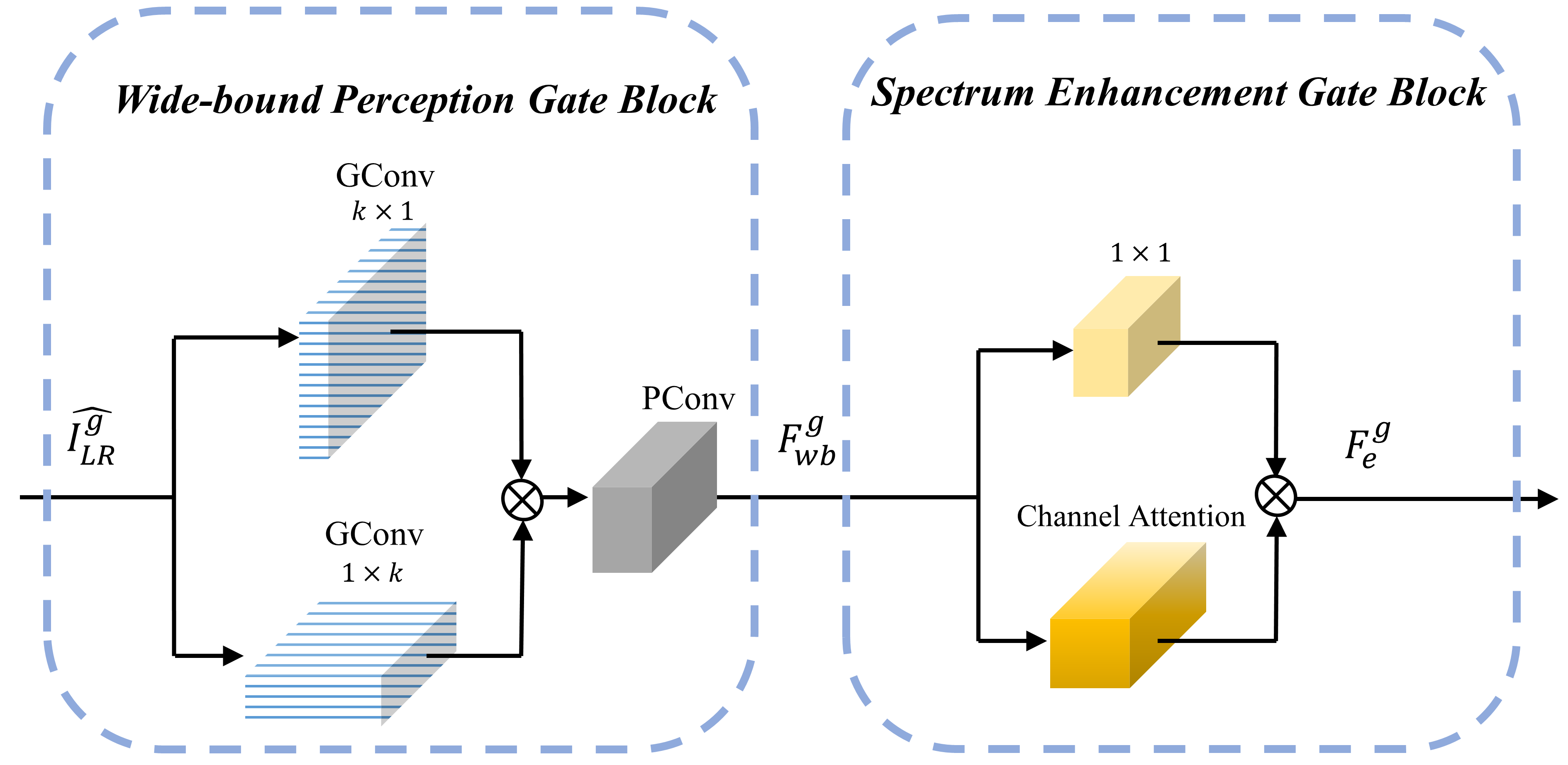}
	\end{subfigure}
	\caption{Structure of the SSRGM.}
	\label{SSRGM}
\end{figure}

Common operations as convolution and self-attention in feature extraction of SHSR always suffer from the inadequate receptive field, high computational cost and spectral distortion. DexiNed \cite{soria2023dense} utilizes multiple depthwise large kernel convolutions to implement the efficient expansion of receptive field. Research from Yang et al. \cite{yang2022focal} reveals element-multiplication gate bring nonlinear and efficient feature extraction. GSSR \cite{wang2023group} demonstrates the concurrent spatial and spectral operation gain the better SHSR. Inspired by the works, we develop a spatial-spectral reinforcement gate module (SSRGM) through employing gate integration of large-kernel strip convolutions and channel mixing attention to extract spatial information efficiently and increase spectral consistency. As shown in the Fig. \ref{SSRGM}, the module consists of the wide-bound perception gate block (WPGB) and spectrum enhancement gate block (SEGB).

Recently, Cordonnier et al. \cite{cordonnier2019relationship} demonstrated that when successive convolutional layers share a common dilation factor, the effective receptive field degenerates into a sparse subset, significantly reducing its coverage. In WPGB, the large kernel group convolution is employed to expand the receptive field. Hu et al. \cite{10453272} suggested that hyperspectral images exhibit axial anisotropy. Thus we decompose the large kernel convolution into two strip convolutions along the horizontal and vertical directions one by one to improve efficiency and capture anisotropic information in hyperspectral images. We further apply the gating mechanism to mutually aggregate the strip convolutions to gain diverse spatial information. Different from the addition operation to linearly merge information, the gate operation can efficiently map diverse features into high-dimensional and nonlinear spaces and enhance the presentation of input features \cite{yang2022focal}.

Then, the partial convolution \cite{10965796}, in which the convolution processes a portion of the channels, is applied to gain wide- bound features $F_{wb}^{g}\in\mathbb{R}^{H \times W \times \frac{C}{G}}$. The abovementioned steps can be formulated as:

\begin{equation}
F_{wb}^{g} = H_{Pc}(H_{Vc}(\widehat{I_{LR}^g})  \ast H_{Hc}(\widehat{I_{LR}^g}))
\end{equation}
Where $H_{Pc}$, $H_{Vc}$ and $H_{Hc}$ denote the functions of the partial convolution and the functions of the vertical and horizontal strip group convolutions.

SEGB employs gating mechanism to dynamically modulate channel attention weights, enabling adaptive calibration of spatial details and spectral significance. This approach simultaneously suppresses inter-channel spectral errors while enhancing the feature representation of spatial structures. The procedure can be formulated as:
\begin{equation}
F_{e}^{g} = H_{CA}(H_{Vc}(F_{wb}^{g})  \ast H_{PW}(F_{wb}^{g})
\end{equation}
Where $H_{CA}$ and $H_{PW}$ denote the functions of the channel attention module and pointwise convolution. $F_{e}^{g}\in\mathbb{R}^{H \times W \times \frac{C}{G}}$ represents the enhanced feature.

\begin{figure}[h!]
	\centering
	\begin{subfigure}{1\linewidth}
		\centering
		\includegraphics[width=0.75\linewidth]{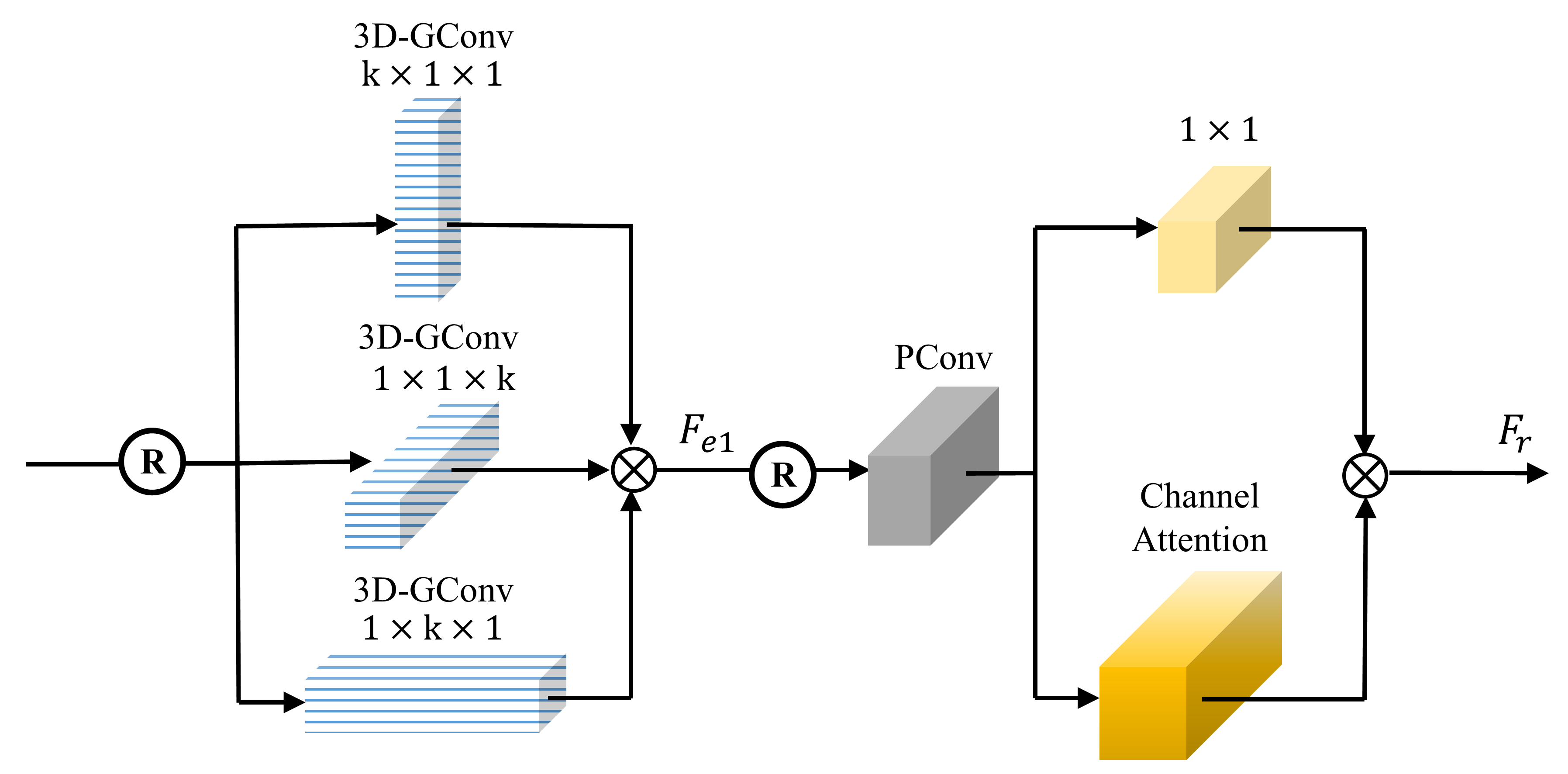}
	\end{subfigure}
	\caption{Structure of the 3D-SSRGM.}
	\label{3DSSRGM}
\end{figure}

As 3D convolution possesses the excellent property of preserving the consistency of the whole feature for high-dimensional data as hyperspectral cube \cite{huang2023spatio}, the SSRGM is upgraded into the 3D-SSRGM to explore spatial and spectral information synchronously in the feature refinement unit as shown in Fig. \ref{3DSSRGM}. We replace original strip convolutions with three strip 3D convolutions in height, width and channel directions to simultaneously refine corresponding feature and enhance geometric perception ability. Furthermore, strip 3D convolution enables extraction of spectral channel anisotropy by building upon spatial anisotropy processing. Compared with other 3D convolution operations, the design can effectively cut down model burden and bring subtle information for each dimension to improve the SR performance. The feature size of this module is properly expanded and squeezed to accommodate the input sharpness of the 2D and 3D convolutions.

\subsection{Loss function}
$L_{1}$ loss is widely used as a loss function in most nature image SR methods to obtain spatial information constraints. This parameter has good convergence and a balanced error distribution. The $L_{1}$ loss is given as follows:
\begin{equation}
L_{1} = \frac{1}{N} \sum_{n=1}^{N} \Vert I_{HR}^{n} - I_{SR}^{n}\Vert_1
\end{equation}
Where $N$ is the batch-size number during the training process. $I_{HR}^{n}$ and $I_{SR}^{n}$ are the $n_{th}$ ground truth hyperspectral image and reconstructed hyperspectral image.

The spectrum loss $L_{spe}$ is proposed to constrain the spectral structure, which can be formulated as:
\begin{equation}
L_{spe} = \frac{1}{N} \sum_{n=1}^{N} \frac{1}{\pi} (\frac{I_{HR}^{n} \cdot I_{SR}^{n}}{\Vert I_{HR}^{n} \Vert_2 \cdot \Vert I_{SR}^{n} \Vert_2) } 
\end{equation}

The gradient information strengthens the SR image sharpness by calculating the difference between neighboring pixels. The gradient map of a hyperspectral image H can be formulated as:
\begin{equation}
\begin{split}
&\nabla H = (\nabla_{h}H, \nabla_{w}H, \nabla_{c}H), \\
&M(H) = \Vert \nabla H \Vert_2
\end{split}
\end{equation}
Where $M(\cdot)$ is the operation used to obtain the gradient map. $\nabla_{h}$, $\nabla_{w}$ and $\nabla_{c}$ denote the horizontal, vertical, and channel gradients of $I_{SR}$. Gradient loss is employed to decrease the distance between different gradient maps of the SR image, which can be formulated as:
\begin{equation}
L_{gra} = \frac{1}{N} \sum_{n=1}^{N} \Vert M(I_{HR}^{n}) - M(I_{SR}^{n})\Vert_1
\end{equation}
In the SSPSR, the spatial–spectral total variation loss $L_{sstv}$ s designed to compensate for context details and concealed information. $L_{sstv}$ can be formulated as:
\begin{equation}
L_{SSVT} = \frac{1}{N} \sum_{n=1}^{N} (\Vert \nabla_{h} I_{SR}^{n}\Vert_1 + \Vert \nabla_{w} I_{SR}^{n}\Vert_1 + \Vert \nabla_{c} I_{SR}^{n}\Vert_1)
\end{equation}
As demonstrated by Wu et al. \cite{wu2023combining}, the combination of various losses promotes SR effectiveness. Similarly, we use the weighted addition of four loss functions as the total loss $L_{total}$, which can be denoted as:
\begin{equation}
L_{total} = L_{1} + \lambda_{1} L_{spe} + \lambda_{2} L_{gra} + \lambda_{3} L_{sstv}
\end{equation}
Where $\lambda_{1}$, $\lambda_{2}$ and $\lambda_{3}$ represents the hyperparameters used to balance the weights of different losses. We follow the pioneer work \cite{9380508}\cite{wang2023spatial} setting $\lambda_{1}$ to 0.5, $\lambda_{2}$ to 0.1 and $\lambda_{3}$ to $1 \times 10^{-3}$ in our experiments.
\section{Experiments and results}
In this section, we reveal implementation details and evaluate our network both qualitatively and quantitatively. The model effectiveness, feature learning range and module necessity are analyzed in efficiency evaluation, receptive field evaluation and ablation study.
\subsection{Implementation details}
We refer to previous work \cite{9380508} in EFGN to group the hyperspectral images into non-overlapping subgroups, each containing $4$ spectral bands. In SPDF, the piece of channel shuffle is set to $4$. In the SSRGM, the input feature map number is set to $64$, the block number is set to $1$ and $6$, corresponding to its normal and 3D versions, and the kernel size of the strip group convolution in the SSRGM is set to $15$. For the training stage, the Adam optimizer is used to train the model over $70$ epochs, and the batch size is set to $16$. The initial learning rate is set to $1 \times 10^{-4}$ and decreases to one-tenth in 30 epochs. The proposed model is executed via PyTorch on an NVIDIA RTX 3090 GPU.
\subsection{Datasets and experimental setup}
The proposed method is evaluated on three common hyperspectral image datasets: the Chikusei dataset, the Pavia Centre dataset, and the Harvard dataset. Our method is compared with six state-of-the-art SHSR models. For fairness of comparison, the hyperparameters of the abovementioned methods are adjusted to achieve their best performance at scale factors of 4 and 8.
Five widely used evaluation metrics are applied to reveal the SR performance, including the peak signal-to-noise ratio (PSNR), structure similarity (SSIM), spectral angle mapper (SAM), root-mean-square error (RMSE), and Erreur Relative Globale Adimensionnelle de Synthèse (ERGAS). These indices gain their best values as $+\infty$, 1, 0, 0, and 0. For the PSNR and SSIM, we calculate the average values of all the spectral bands. SAMs are commonly used to evaluate spectral fidelity. The others usually estimate the reconstruction effect of spatial information.
Three evaluation indices are adopted to verify the network efficiency, including parameters (Params), floating point operations per second (FLOPs), and inference memory (Memory). Params represents the parameters of the whole network, and FLOPs is utilized to measure the computational complexity. Memory refers to the resource occupancy required to complete the SR procedure.
\subsection{Experimental results on the Chikusei dataset}
\begin{figure}[h]
	\centering
	\begin{subfigure}{1\linewidth}
		\centering
		\includegraphics[width=1\linewidth]{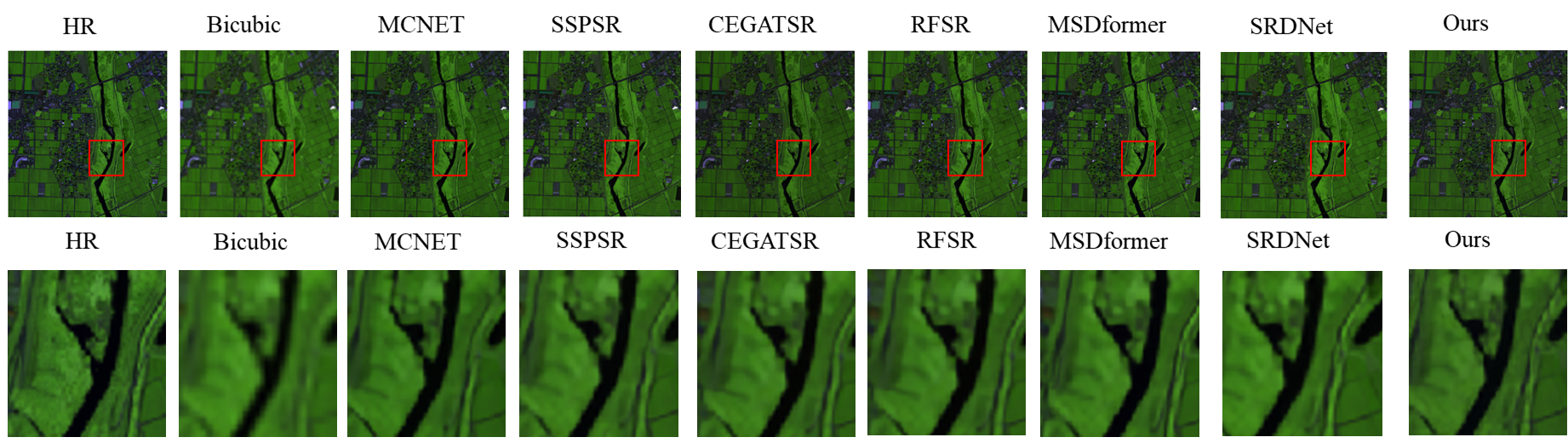}
	\end{subfigure}
	\caption{Reconstructed images of test hyperspectral images in the Chikusei dataset with spectral bands 70–100–36 as R–G–B with the scale factor of 4. Left to right: Ground truth and the results of Bicubic, MCNET, SSPSR, CEGATSR, RFSR, MSDformer, SRDNet, and our method.}
	\label{chikusei1}
\end{figure}

The Chikusei dataset is a remote sensor hyperspectral image dataset taken by the Headwall Hyperspect-VNIR-C imaging sensor covering agricultural and urban areas in Chikusei, Ibaraki, Japan. This dataset consists of 2517×2335 pixels and has 128 spectral bands from the spectral range of 363 nm to 1018 nm. We cut the center region of the entire image for testing and training. The top region of this image is cropped into 4 nonoverlapping images with 512×512×128 pixels as testing data. The remaining part is extracted to form overlap patches for training data ($10\%$ is used for validation). The large-scale factor needs more input information to stabilize the training process. We extract patches with a size of 64×64 pixels with 32 overlapping pixels and 128×128 pixels with 64 overlapping pixels for the experiments with scale factors of 4 and 8, respectively.
Table \ref{chikusei} shows the comparison of the proposed method with several state-of-the-art methods on the Chikusei testing set. The best and second-best results are bolded and underlined. The EFGN achieves the best performance with respect to all objective evaluation indices. The PSNR is approximately 0.15 and 0.32 higher, and the SAM is 0.12 and 0.2 lower than the CEGATSR at ×4 and ×8 scale factors. Although MCNET alternately applies 2D and 3D convolutions to explore holistic features to some degree, the SR performance is still insufficient. Benefiting from the dynamic weighting, SRDNet adaptively gains learnable message to obtain acceptable performance. With the group strategy, the SSPSR achieves suboptimal SR results at a ×8 scale factor. With the same group pattern, CEGATSR achieves commendable performance at a ×4 scale factor because of the flexible weight allocation given by the graph attention mechanism.
\begin{table*}[t]
	\tiny
	\centering
	
	\resizebox{0.77\linewidth}{!}{
		\begin{tabu}{lcccccc}
			\hline
			\specialrule{0em}{1pt}{0pt}
			{Scale} &{Method}  & {PSNR$\uparrow$}  & {SSIM$\uparrow$} &{SAM$\downarrow$} & {RMSE$\downarrow$}& {ERGAS$\downarrow$} \\
			\specialrule{0em}{1pt}{0pt}
			\hline

		\multirow{8}{*}{$\times 4$} &Bicubic  &37.6377	&0.8953	&3.4039	&0.0155	&6.7563                                            \\
			&MCNET    &39.4337	&0.9300	&2.8869	&0.0129	&5.4280        \\
			&SSPSR  &39.9815	&0.9391	&2.4781	&0.0118	&5.1725                          \\
			&CEGATSR   &\underline{40.2206}	&\underline{0.9424}	&\underline{2.4092}	&\underline{0.0115}	&\underline{5.0301}                         \\
			&RFSR   &39.9469	&0.9387	&2.6260	&0.0119	&5.2184                        \\
			&MSDFormer &39.6497	&0.9366	&2.4443	&0.0123	&5.3603  \\
			&SRDNet  &40.1236	&0.9414	&2.4061	&0.0116	&5.1026	
			\\

			&EFGN(Ours) &$\textbf{40.3953}$	&$\textbf{0.9451}$	&$\textbf{2.2702}$ &$\textbf{0.0113}$	&$\textbf{4.8531}$
			\\
			
			\specialrule{0em}{1pt}{0pt}
			\hline
			\specialrule{0em}{2pt}{0pt}
			\multirow{8}{*}{$\times 8$} &Bicubic   &34.5048	&0.8068	&5.0435	&0.0223	&9.6975            \\
		    &MCNET    &35.4231	&0.8365	&4.5097	&0.0202	&8.6754              \\
			&SSPSR  &\underline{35.5615}	&\underline{0.8446}	&\underline{4.2067}	&\underline{0.0197}	&\underline{8.5894}                            \\
			&CEGATSR  &35.5354	&0.8444	&4.2166	&0.0197	&8.6269                          \\
			&RFSR &33.9411	&0.7843	&5.3647	&0.0241	&10.2277                     \\
			&MSDFormer &35.3729	&0.8395	&4.3218	&0.0203	&8.6807  \\
			&SRDNet  &35.5362	&0.8386	&4.2365	&0.0236	&8.6345	
			\\
			&EFGN(Ours)    &$\textbf{35.8887}$	&$\textbf{0.8558}$	&$\textbf{4.0080}$	&$\textbf{0.0189}$	&$\textbf{8.2932}$
			\\
			
			\specialrule{0em}{1pt}{0pt}
			\hline
			
		\end{tabu}
	}
	\caption{Quantitative evaluation on the Chikusei dataset of state-of-the-art hyperspectral image SR algorithms. The best results are shown in $\textbf{Bold}$ and the second best results are shown in $\text{\underline{Underline}}$.}
	\label{chikusei}
\end{table*}
The reconstructed results are compared intuitively in this study. In particular, we visualize the qualitative results of the SR image from the Chikusei test set at the ×4 scale factor by selecting the 70th, 100th, and 36th bands as the R–G–B channel to form the pseudo-RGB image, and the clearly contrasting area is enlarged in Fig. \ref{chikusei1}. The bicubic result is blurred over the whole area. MCNET and RFSR produce blurred edges and details in SR images. Despite obtaining fair reconstruction results, the SSPSR and CEGATSR methods still have shortcomings in restoring contour lines. The image edges of MSDformer and SRDNet are extremely sharp and seem unreal. Our method obtains the best spatial fidelity in image contours or details.

\begin{figure*}[h]
	\centering
	\begin{subfigure}{1\linewidth}
		\centering
		\includegraphics[width=1\linewidth]{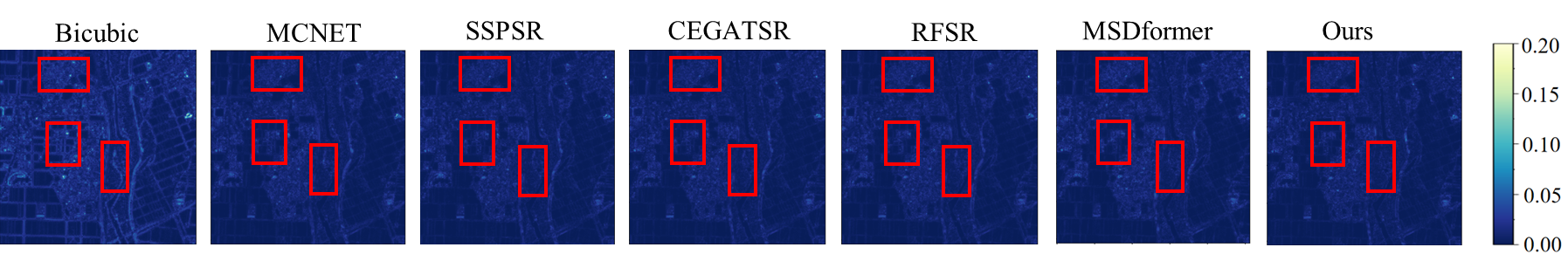}
	\end{subfigure}
	\caption{Mean error maps of test hyperspectral images in the Chikusei dataset at the scale factor of 4. Left to right: results of Bicubic, MCNET, SSPSR, CEGATSR, RFSR, MSDformer, SRDNet, and our method.}
	\label{chikusei2}
\end{figure*}

\begin{figure*}[h]
	\centering
	\begin{subfigure}{0.48\linewidth}
		\includegraphics[width=\linewidth]{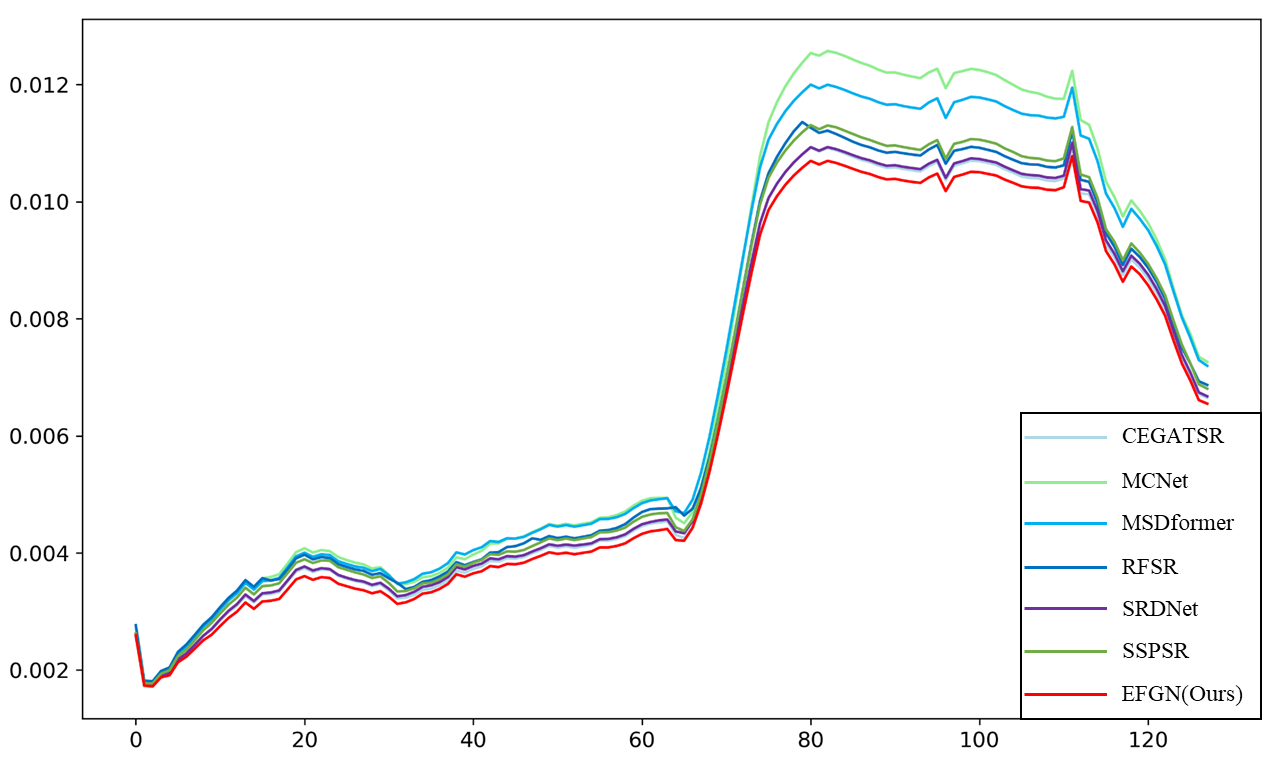}
	\end{subfigure}
	\hfill 
	\begin{subfigure}{0.48\linewidth} 
		\includegraphics[width=\linewidth]{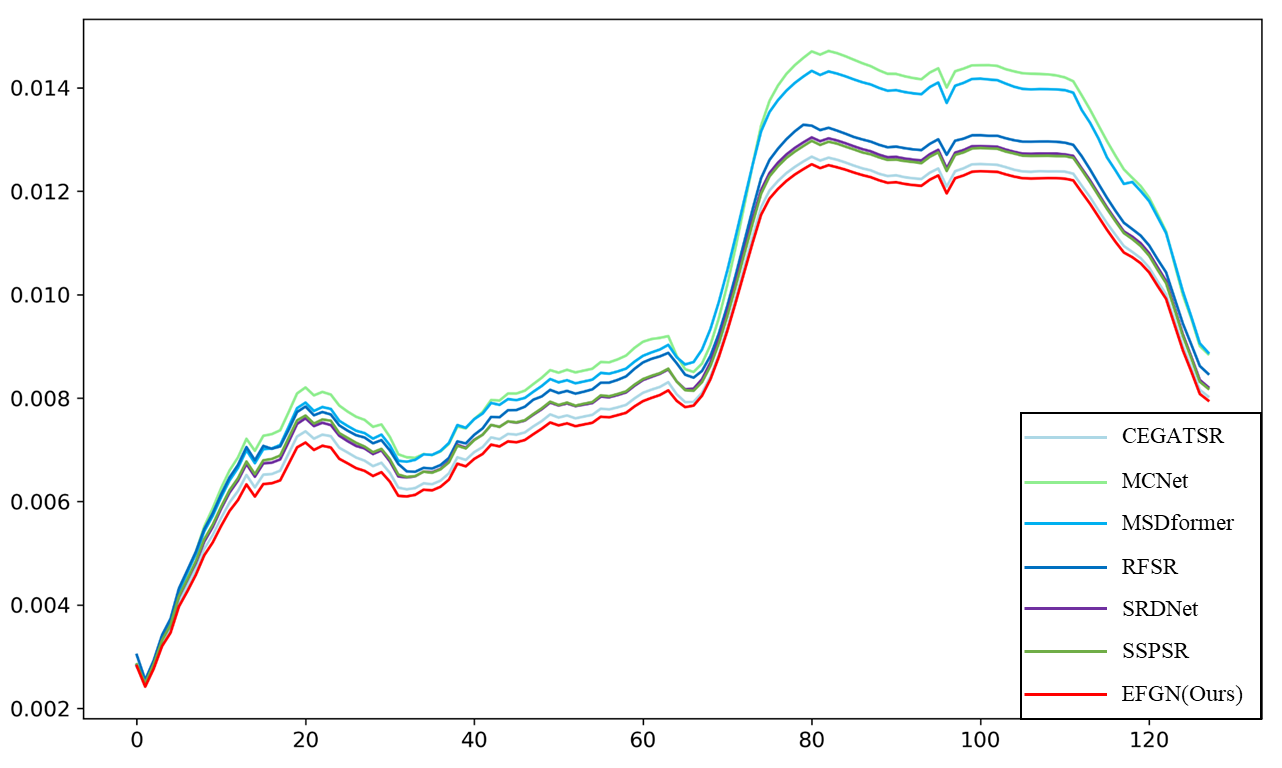}
	\end{subfigure}
	\caption{Mean spectral difference curve of two test hyperspectral images in the Chikusei dataset at the scale factor of 4.}
	\label{chikusei3}
\end{figure*}

We visualize the spatial mean error between the SR and HR images in Fig. \ref{chikusei2}. The brighter the image is, the lower the accuracy of the SR reconstruction. Bicubic possesses the lightest pixel zone in the whole error map, and others produce bluer pixel regions but still retain the light spot in particular areas. The EFGN provides the least number of bright pixels in the selected area and demonstrates superior reconstruction ability.
The mean spectral difference curve of two test images is presented in Fig. \ref{chikusei3}. As the bands increase, the spectral difference gradually rises and reaches the maximum at the 70th band. Then the spectral difference descends along with the ascension of bands. Bicubic obtains the largest difference in most bands while CEGATSR and RFSR reduce the spectral difference obviously, but our method still possesses the lowest reconstruction difference curve in all bands.

\subsection{Experimental results on the Pavia dataset}
The Pavia Centre dataset was obtained via a reflective optical system imaging spectrometer sensor during a flight campaign over Pavia in northern Italy in 2001. We select the center part of the scene and remove some noisy bands to form the final image with 1096×715×102 pixels in total. Our method is evaluated in this study. In particular, we crop the left region of the image into four 223×223×102 pixels images as testing data. The remaining region is cropped to form the overlapping patches for training. The patches are cropped to a size of 64×64 pixels with 32 overlapping pixels and 128×128 pixels with 64 overlapping pixels at scale factors of 4 and 8.

Table \ref{pavia} shows the quantitative results of all methods in terms of five evaluation metrics on the Pavia data testing set at scale factors of 4 and 8. Given the limited number of training samples, the results of each method on this dataset are much lower than those on the other datasets. The proposed EFGN achieves the best performance at scale factors of 4 and 8. The PSNR and SAM are 0.16 higher and 0.25 lower than the second-best results. According to Yu [30], the large kernel convolutions and their variants possess strong feature collection ability, which can obtain more information with limited learning resources. This situation may explain why the EFGN achieves great performance on this small-sample dataset.

\begin{table*}[t]
	\tiny
	\centering
	
	\resizebox{0.77\linewidth}{!}{
		\begin{tabu}{lcccccc}
			\hline
			\specialrule{0em}{1pt}{0pt}
			{Scale} &{Method}  & {PSNR$\uparrow$}  & {SSIM$\uparrow$} &{SAM$\downarrow$} & {RMSE$\downarrow$}& {ERGAS$\downarrow$} \\
			\specialrule{0em}{1pt}{0pt}
			\hline

			\multirow{8}{*}{$\times 4$} &Bicubic  &27.5874	&0.7217	&6.1398	&0.0437	&6.8813              \\
			&MCNET    &28.8155	&0.8013	&5.9976	&0.0378	&5.9895      \\
			&SSPSR    &28.9239  &0.8050	&5.7029	&0.0372	&5.9264                          \\
			&CEGATSR   &28.9970	&0.8082	&5.6703	&0.0368	&5.8868                         \\
			&RFSR   &\underline{29.0269}	&0.8091	&5.8484	&\underline{0.0367}	&\underline{5.8654}                        \\
			&MSDFormer &28.9906	&\underline{0.8120}	&5.6168	&0.0369	&5.8852  \\
			&SRDNet  &28.8806	&0.8053	&\underline{5.4997}	&0.0374	&5.9481	
			\\
			&EFGN(Ours) &$\textbf{29.1756}$	&$\textbf{0.8159}$	&$\textbf{5.3821}$ &$\textbf{0.0361}$	&$\textbf{5.7603}$
			\\
			
			\specialrule{0em}{1pt}{0pt}
			\hline
			\specialrule{0em}{2pt}{0pt}
			\multirow{8}{*}{$\times 8$} &Bicubic   &24.5972	&0.4590	&7.8478	&0.0629	&9.6569 \\
			&MCNET    &25.0312	&0.5137	&\underline{7.7421}	&0.0598	&9.1843              \\
			&SSPSR  &25.0511	&0.5169	&7.7477	&0.0595	&9.1677                            \\
			&CEGATSR  &24.9987	&0.5107	&7.8327	&0.0599	&9.2230                          \\
			&RFSR &\underline{25.0577}	&0.5138	&7.7454	&\underline{0.0595}	&\underline{9.1580}                     \\
			&MSDFormer &25.0031	&\underline{0.5174}	&7.7568	&0.0598	&9.2244  \\
			&SRDNet &25.0465	&0.5135	&7.7521	&0.0597	&9.2065
			\\
			&EFGN(Ours)    &$\textbf{25.1352}$	&$\textbf{0.5282}$	&$\textbf{7.7301}$	&$\textbf{0.0589}$	&$\textbf{9.0819}$
			\\
			
			\specialrule{0em}{1pt}{0pt}
			\hline
			
		\end{tabu}
	}
	\caption{Quantitative evaluation on the Pavia dataset of state-of-the-art hyperspectral image SR algorithms. The best results are shown in $\textbf{Bold}$ and the second best results are shown in $\text{\underline{Underline}}$.}
	\label{pavia}
\end{table*}

\begin{figure}[h!]
	\centering
	\begin{subfigure}{1\linewidth}
		\centering
		\includegraphics[width=1\linewidth]{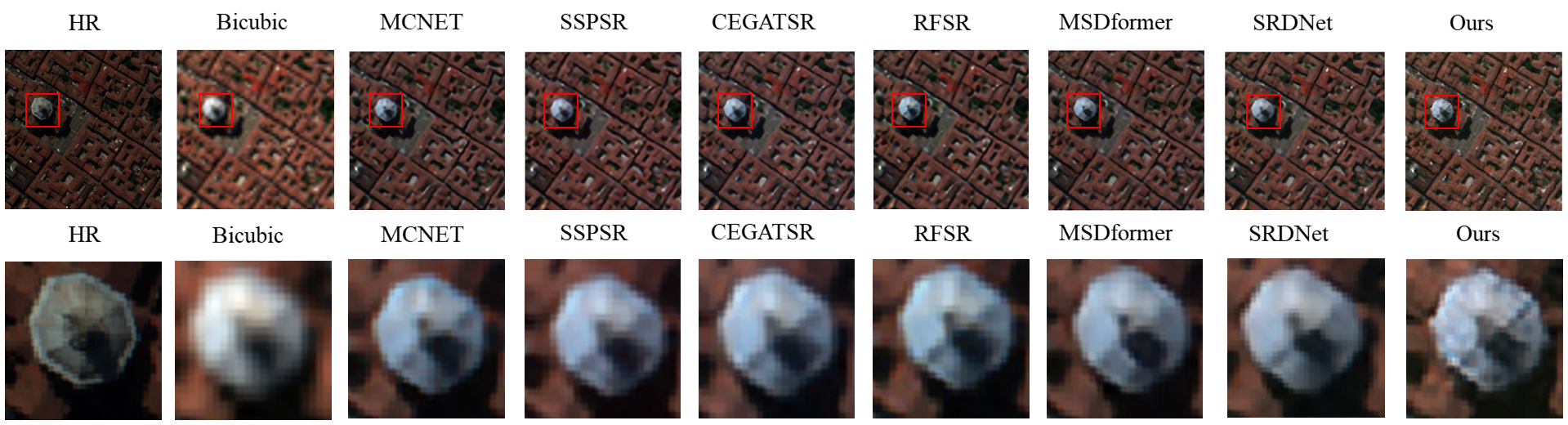}
	\end{subfigure}
	\caption{Reconstructed images of test hyperspectral images in the Pavia Centre dataset with spectral bands 60–31–12 as R–G–B with the scale factor of 4. Left to right: ground truth and the results of Bicubic, MCNET, SSPSR, CEGATSR, RFSR, MSDformer, SRDNet, and our method.}
	\label{pavia1}
\end{figure}

\begin{figure}[h!]
	\centering
	\begin{subfigure}{1\linewidth}
		\centering
		\includegraphics[width=1\linewidth]{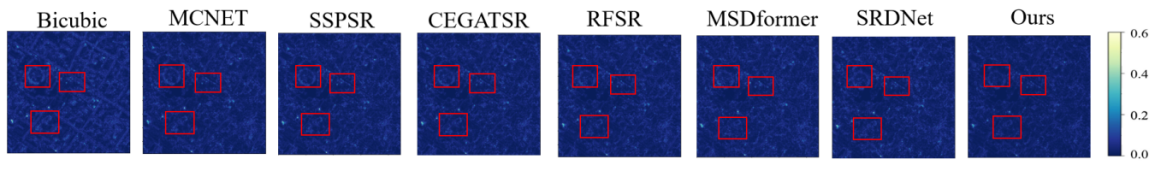}
	\end{subfigure}
	\caption{Mean error maps of test hyperspectral images in the Pavia dataset at the scale factor of 4. Left to right: results of Bicubic, MCNET, SSPSR, CEGATSR, RFSR, MSDformer, SRDNet, and our method.}
	\label{pavia2}
\end{figure}

\begin{figure*}[h]
	\centering
	\begin{subfigure}{0.48\linewidth} 
		\includegraphics[width=\linewidth]{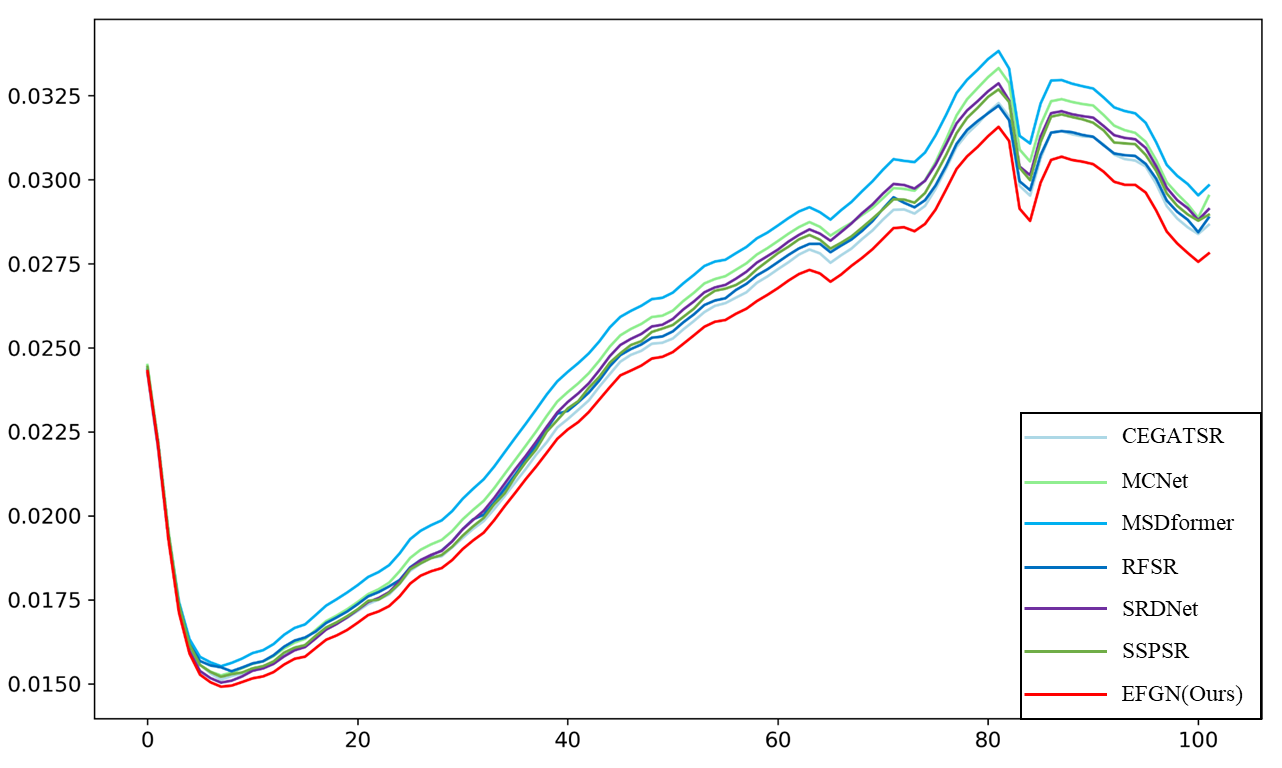}
	\end{subfigure}
	\hfill 
	\begin{subfigure}{0.48\linewidth} 
		\includegraphics[width=\linewidth]{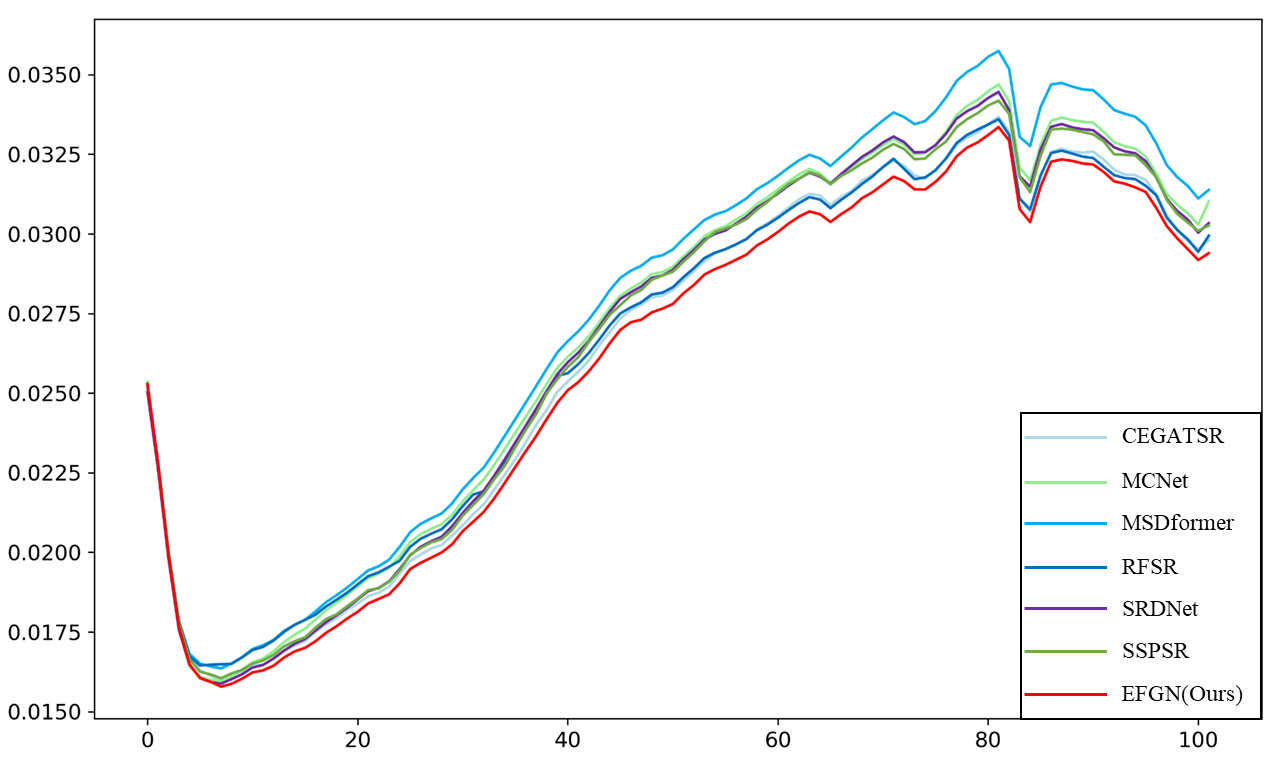}
	\end{subfigure}
	\caption{Mean spectral difference curve of two test hyperspectral images in the Pavia dataset at the scale factor of 4.}
	\label{pavia3}
\end{figure*}

Fig. \ref{pavia1} shows the pseudo-RGB images. These images can be visualized by treating the 60th, 31st, and 12th bands as R–G–B channels from the Pavia test set at the scale factor of 4. The results of MCNET and SSPSR seem to be fuzzy. MSDformer and RFSR obtain better reconstruction results but unsharp boundaries. The result of SRDNet seems to be smooth but still blurry. Compared with other methods, our method recovers clearer texture lines and fewer artifacts. The mean error maps are shown in Fig. \ref{pavia2}. Bicubic and MCNET provide lighter spots in the street and building margins, and the EFGN possesses the bluest area in the object contour and detail. From the mean spectral difference curve of the Pavia test images at the scale factor of 4 in Fig. \ref{pavia3}, our method provides the lowest error in most spectral bands.

\subsection{Experimental results on the Harvard dataset}
The Harvard dataset is obtained via a Nuance FX, CRI INC camera. This dataset consists of 50 hyperspectral images of real-world indoor and outdoor scenes in daylight. Each image contains 1024×1392 pixels with 31 spectral bands ranging from 420 to 720 nm. To prepare the training samples, we randomly pick 30 hyperspectral images and extract patches of 64×64 pixels and 128×128 pixels at scale factors of 4 and 8. The remaining 20 images are cropped into patches of 512×512 pixels as testing samples.
The performance of the different methods at scale factors of 4 and 8 in terms of the five evaluation metrics is shown in Table \ref{harvard}. For the ×8 scale factor, RFSR possesses the best spectral fidelity, which can be attributed mainly to the regularization module it proposed. Compared with the other methods, the proposed EFGN achieves the best SR results for ×4 and ×8 scale factors. The Harvard dataset, which contains images with different scenes and proves the robustness and generalizability of our method.

\begin{table*}[t]
	\tiny
	\centering
	
	\resizebox{0.77\linewidth}{!}{
		\begin{tabu}{lcccccc}
			\hline
			\specialrule{0em}{1pt}{0pt}
			{Scale} &{Method}  & {PSNR$\uparrow$}  & {SSIM$\uparrow$} &{SAM$\downarrow$} & {RMSE$\downarrow$}& {ERGAS$\downarrow$} \\
			\specialrule{0em}{1pt}{0pt}
			\hline

			\multirow{8}{*}{$\times 4$} &Bicubic  &64.3495	&0.99882	&3.2016	&0.00083	&4.7035\\
			&MCNET    &\underline{65.6788}	&\underline{0.99916}	&\underline{3.0913}	&\underline{0.00071}	&\underline{4.0572}       \\
			&SSPSR    &65.6108	&0.99915 &3.0939 &0.00072&4.0999                        \\
			&CEGATSR   &65.5252	&0.99913	&3.0976	&0.00073	&4.1361                    \\
			&RFSR   &65.0555	&0.99913	&3.6150	&0.00077	&4.5514                       \\
			&MSDFormer &65.4574	&0.99911	&3.1522	&0.00075	&4.1666  \\
			&SRDNet  &65.5746	&0.99911	&3.1623	&0.00073	&4.1064	
			\\
			&EFGN(Ours) &$\textbf{65.7060}$	&$\textbf{0.99917}$	&$\textbf{3.0916}$ &$\textbf{0.00070}$	&$\textbf{4.0471}$
			\\
			
			\specialrule{0em}{1pt}{0pt}
			\hline
			\specialrule{0em}{2pt}{0pt}
			\multirow{8}{*}{$\times 8$} &Bicubic   &60.9511	&0.99681 &3.7224 &0.00136	&6.8271 \\
			&MCNET   &\underline{61.9515}	&\underline{0.99762}	&\underline{3.6167}	&\underline{0.00114}	&\underline{6.1594}              \\
			&SSPSR  &61.9230	&0.99762	&3.6267	&0.00115	&6.3569                       \\
			&CEGATSR &61.3439	&0.99712	&3.8332	&0.00122	&6.5415                             \\
			&RFSR &61.9465	&0.99761	&3.6599	&0.00118	&6.1601                    \\
			&MSDFormer &61.2293	&0.99711	&3.6956	&0.00127	&6.6087  \\
			&SRDNet &61.9432	&0.99751	&3.6533	&0.00126	&6.4362
			\\
			&EFGN(Ours)    &$\textbf{61.9586}$	&$\textbf{0.99765}$	&$\textbf{3.6123}$	&$\textbf{0.00112}$	&$\textbf{6.1556}$
			\\
			
			\specialrule{0em}{1pt}{0pt}
			\hline
			
		\end{tabu}
	}
	\caption{Quantitative evaluation on the Harvard dataset of state-of-the-art hyperspectral image SR algorithms. The best results are shown in $\textbf{Bold}$ and the second best results are shown in $\text{\underline{Underline}}$.}
	\label{harvard}
\end{table*}

\begin{figure}[h!]
	\centering
	\begin{subfigure}{1\linewidth}
		\centering
		\includegraphics[width=1\linewidth]{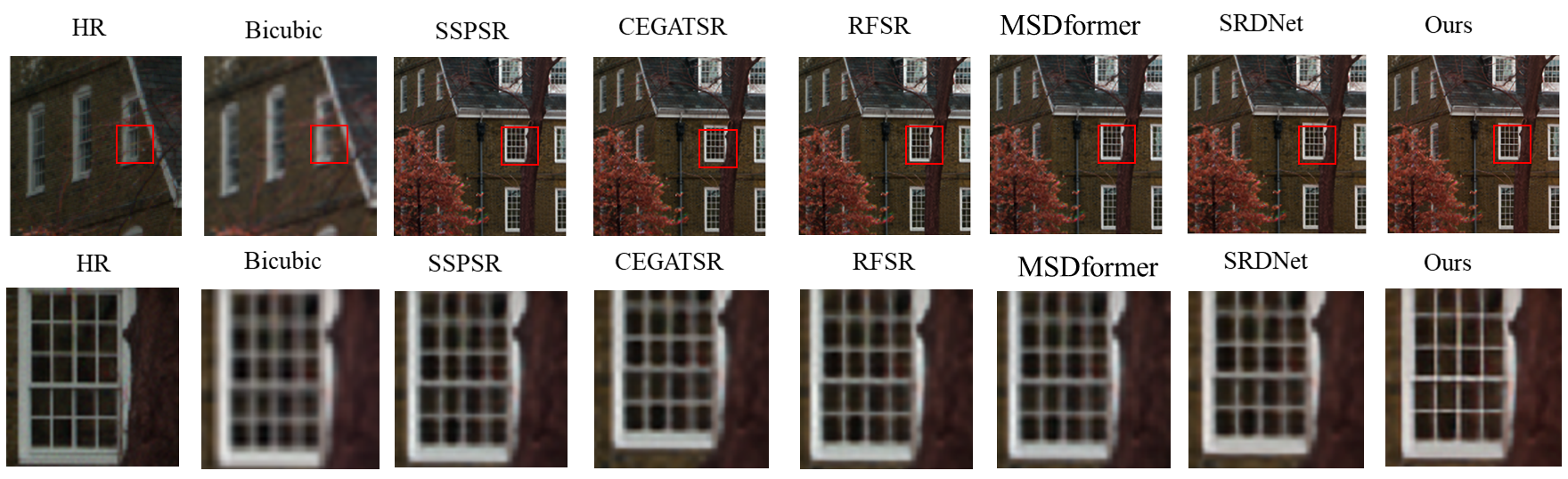}
	\end{subfigure}
	\caption{Reconstructed images of test hyperspectral images in the Harvard dataset with spectral bands 31–20–15 as R–G–B with the scale factor of 4. Left to right: ground truth and the results of Bicubic, MCNET, SSPSR, CEGATSR, RFSR, MSDformer, SRDNet, and our method.}
	\label{harvard1}
\end{figure}

\begin{figure}[h!]
	\centering
	\begin{subfigure}{1\linewidth}
		\centering
		\includegraphics[width=1\linewidth]{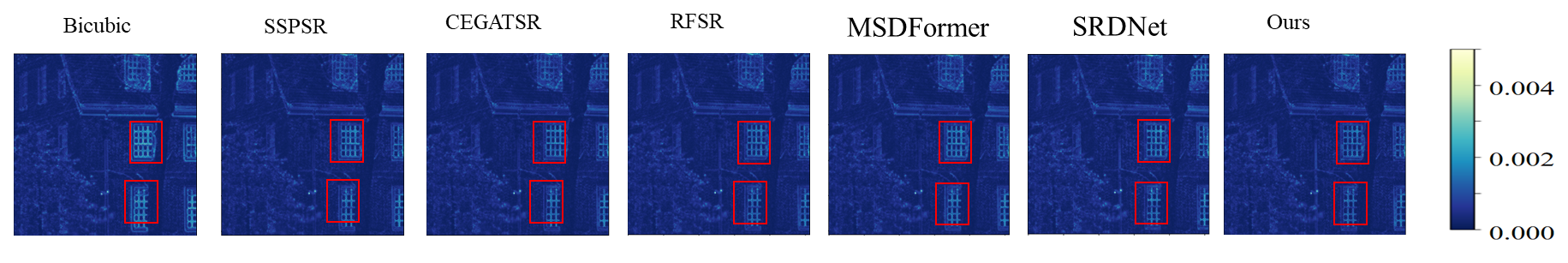}
	\end{subfigure}
	\caption{Mean error maps of test hyperspectral images in the Harvard dataset at the scale factor of 4. Left to right: results of Bicubic, MCNET, SSPSR, CEGATSR, RFSR, MSDFormer, SRDNet, and our method.}
	\label{harvard2}
\end{figure}

\begin{figure*}[h]
	\centering
	\begin{subfigure}{0.48\linewidth} 
		\includegraphics[width=\linewidth]{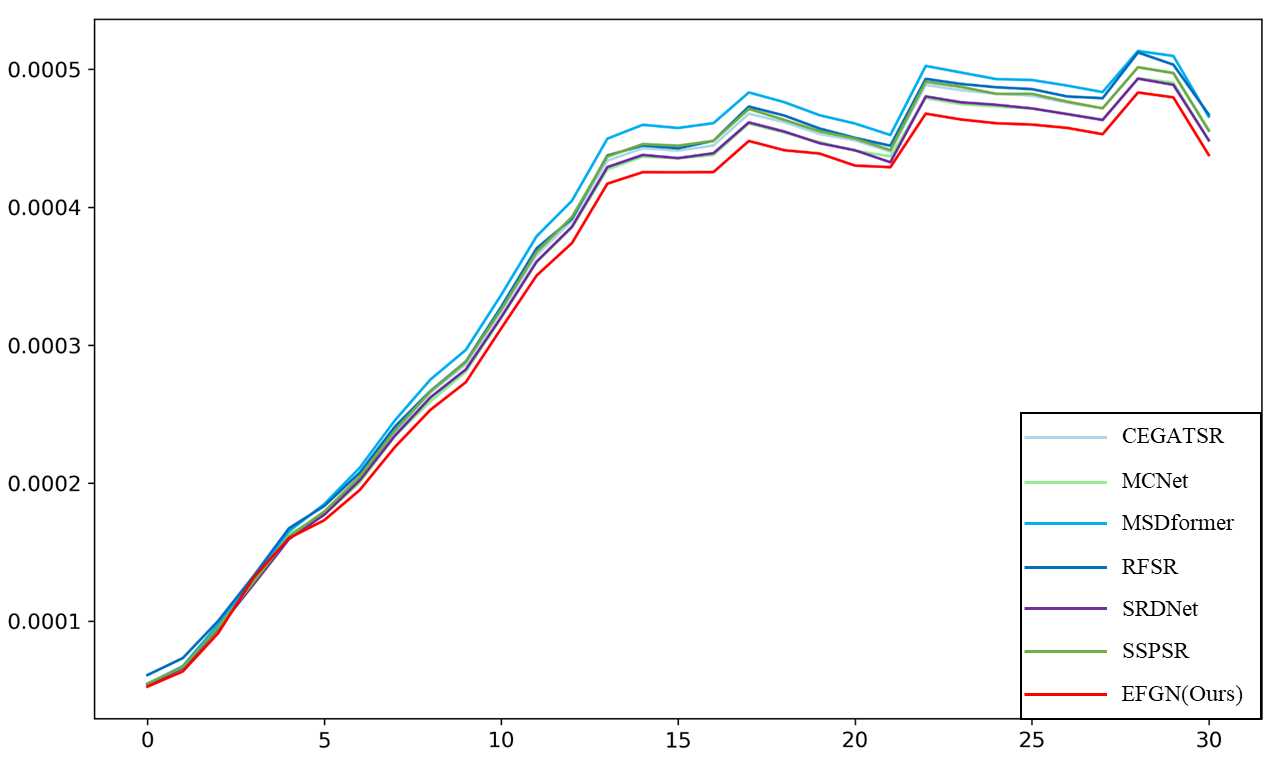}
	\end{subfigure}
	\hfill
	\begin{subfigure}{0.48\linewidth}
		\includegraphics[width=\linewidth]{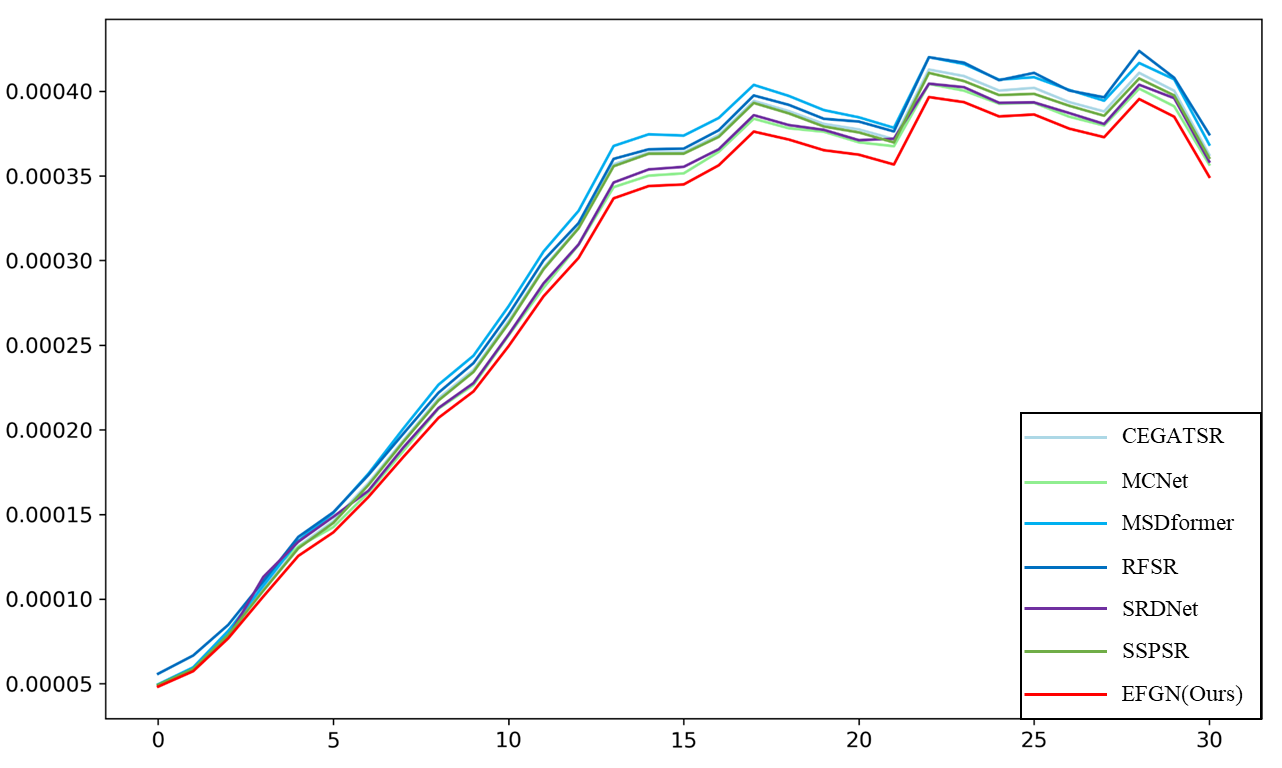}
	\end{subfigure}
	\caption{Mean spectral difference curve of two test hyperspectral images in the Harvard dataset at the scale factor of 4.}
	\label{harvard3}
\end{figure*}

We visualize the pseudo-RGB image by selecting the 31st, 20th, and 15th bands as the R-G-B channels of the testing dataset in Fig. \ref{harvard1}. CEGATSR produces many artifacts, whereas RFSR presents an oversharp margin that is unreal. Our method can restore clearer images than the other methods. From the visualization results of the mean error map in Fig. \ref{harvard2}, the EFGN is generally darker, especially in the lower building window, which reveals a better recovery effect. Fig. \ref{harvard3} shows the mean error curves of all methods for two test images. The error constantly increases as the number of bands increases, which indicates that the redshift band is more difficult to reconstruct, and our method achieves the minimum reconstruction error in most spectral bands.

\subsection{Efficiency evaluation}
A model burden comparison of all methods is shown in Table \ref{efficiency}. Params, FLOPs, and Memory are calculated on the Chikusei dataset at the scale factor of ×4. The Params of the SSPSR and MSDformer are enormous and ten times larger than those of the EFGN. The Memory of MCNET and CEGATSR are unbearable and increase the model burden. The proposed EFGN yields the lowest FLOPs and Memory and the second lowest Params.
\begin{table*}[t]
	\tiny
	\centering
	
	\resizebox{0.77\linewidth}{!}{
		\begin{tabu}{lcccccc}
			\hline
			\specialrule{0em}{1pt}{0pt}
			{Method}  & {}  & {Params(M)} &{} & {FLOPs(G)}& {} &{Memory(G)}\\
			\specialrule{0em}{1pt}{0pt}
			\hline

			MCNET    & &2.17	& &465.60	& &53.98    \\
			SSPSR    & &13.56	& &676.48	& &4.37                        \\
			CEGATSR   & &1.19	& &67.44	& &26.73                   \\
			RFSR   & &0.98	& &51.10	& &5.24                   \\
			MSDFormer & &15.53	& &183.94	& &4.46 \\
			SRDNet  & &1.78	& &59.67	& &9.25
			\\
			EFGN(Ours) & &1.13	& &50.02	& &3.06
			\\
			
			\specialrule{0em}{1pt}{0pt}
			\hline
			
		\end{tabu}
	}
	\caption{Model burden comparison on the Chikusei dataset at ×4 scale factors of state-of-the-art hyperspectral image SR algorithms.}
	\label{efficiency}
\end{table*}

\begin{figure}[h!]
	\centering
	\begin{subfigure}{1\linewidth}
		\centering
		\includegraphics[width=1\linewidth]{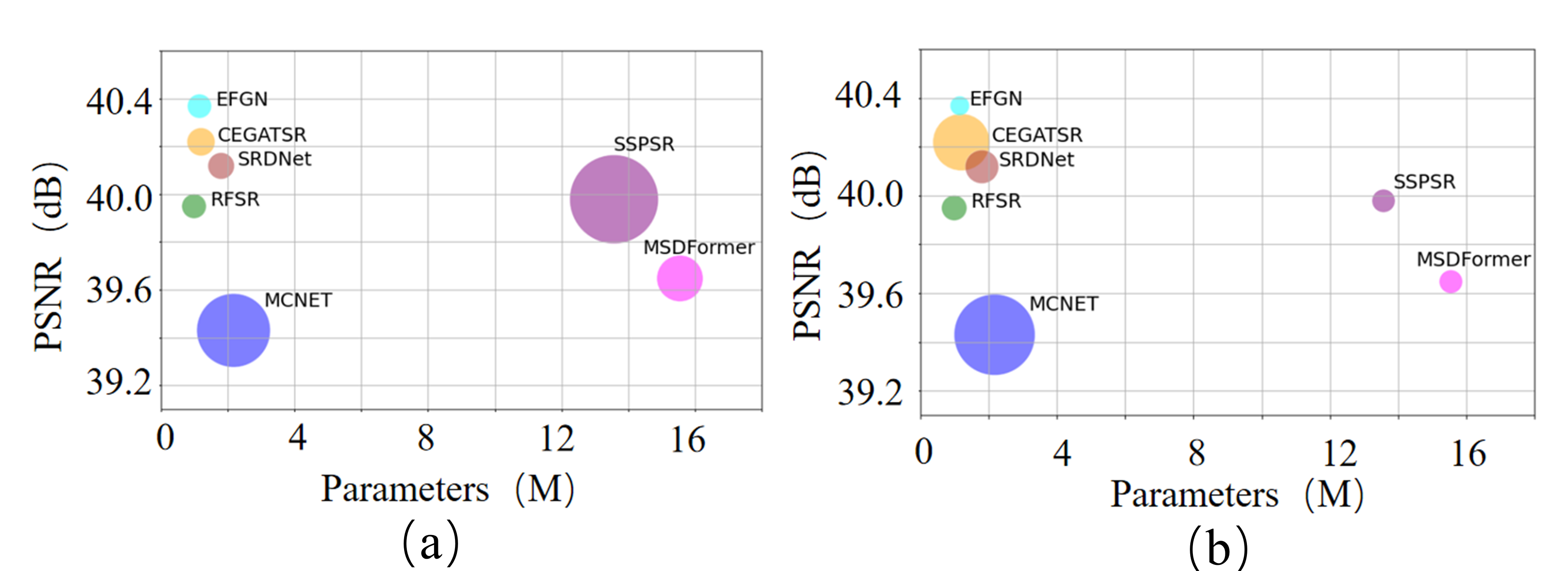}
	\end{subfigure}
	\caption{Visualization of model burden comparison. (a) PSNR vs. Parameters and FLOPs, (b) PSNR vs. Parameters and Memory.}
	\label{efficiency1}
\end{figure}

The performance, Params, FLOPs, and Memory of the GPU are determined to intuitively illustrate the efficiency of our method in Fig. \ref{efficiency1}. The horizontal and vertical coordinates represent the model parameters and PSNR. The circles indicate the FLOPs and Memory respectively. As 3D convolution is utilized, MCNET possesses more FLOPs and Memory than the other methods, and the SR performance is poor. The high computational complexity and inference resource occupancy limit its usage. The SSPSR and MSDformer achieve mediocre reconstruction performance with large Params. Although CEGATSR and RFSR have low Params, FLOPs, and Memory, the reconstruction results are worse than those of our method. The proposed EFGN achieves the best SR performance with the lowest model burden.

\subsection{Receptive field evaluation}
\begin{figure}[t]
	\centering
	\begin{subfigure}{1\linewidth}
		\centering
		\includegraphics[width=1\linewidth]{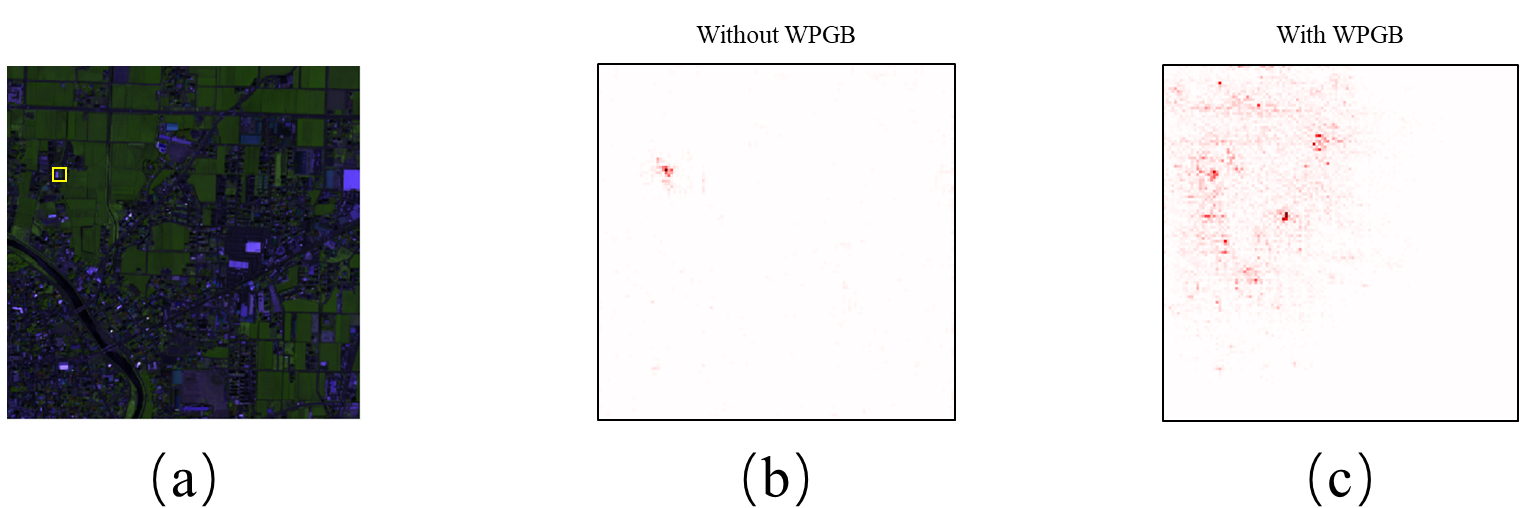}
	\end{subfigure}
	\caption{Comparison of receptive field on a Chikusei test image (×4) using contribution map. The pseudo-RGB image of the test image (a); EFGN without WPGB(b) EFGN with WPGB(c)}
	\label{lam}
\end{figure}

In order to further validate the effectiveness of WPGB in obtaining contextual information, we use the local attribution map \cite{wang2025asymmetric} to design a visualization experiment as shown in Fig. \ref{lam} to demonstrate the specific impact of this module on the receptive field. The red region in the figure represents the utilized context region, and the higher saturation of red color means the deeper utilization. Visual comparison reveals a significant expansion and intensification of the red region in the rightmost figure compared to the center image. This enhancement demonstrates the efficacy of the Weighted Progressive Gating Block (WPGB) in expanding the effective receptive field for the Enhanced Feature Guidance Network (EFGN). By strategically integrating multi-directional large-kernel strip convolutions through adaptive gating mechanisms, WPGB enables comprehensive cross-boundary feature fusion. This architectural approach facilitates the aggregation of rich contextual information across spatial dimensions, thereby maximizing the network's perceptual scope while maintaining feature coherence.

\subsection{Effectiveness evaluation}

\begin{itemize}
	\item[$\bullet$]Effectiveness of the SSRGM
	
	Ablation experiments are conducted on the WPGB and SEGB to verify the effectiveness of each block in the SSRGM in Table \ref{ablation}. When only the WPGB is utilized, the reconstruction performance is inferior. When SEGB is adopted, the PSNR increases by 0.03, and the SAM decreases by 0.01. This trend proves that band consistency is beneficial for the SHSR. When WPGB and SEGB are employed, the PSNR increases by 0.03, whereas SAM and ERGAS decrease by 0.02 and 0.01, respectively. This finding demonstrates that two blocks are indispensable in the SSRGM to extract spatial–spectral information better.
	\item[$\bullet$]Effectiveness of the SPDFM
	
	The effectiveness of the SPDFM is evaluated in Table \ref{ablation}. SC denotes spectral channel guidance, and  represents hierarchical spatial guidance. With spectral channel guidance, the PSNR increases by 0.03, and the SAM decreases by 0.04. The results show that primary and adjacent information is favorable for SRs. With hierarchical spatial guidance, improvements in reconstruction are limited. When both SC and HS are used, the PSNR, SSIM, SAM, RMSE, and ERGAS are 0.05 higher, 0.003 higher, 0.07 lower, 0.0003 lower, and 0.04 lower, than the results without the SPDFM. According to Wang et al. \cite{wang2023remote}, multilevel features can enhance network SR capability and maintain the integrity of extracted information. Multiple types of assistance are assumed to mutually bring complementary information to improve SR performance.
	\item[$\bullet$]Effectiveness of the 3D-SSRGM
	
	When we use the 3D-SSRGM after subgroups merge (represented by “3D” in Table \ref{ablation}), the PSNR and SSIM increase by 0.04 and 0.001, whereas the SAM, RMSE, and ERGAS decrease by 0.02, 0.0002, and 0.03. The spatial fidelity increases, and the spectral distortion decreases. According to Azad et al. \cite{azad2024beyond}, 3D large-kernel convolution surpass the 2D convolution by enabling cross-band parameter sharing, which more effectively models spatial-spectral joint representations. This approach simultaneously enhances spatial feature representation while significantly improving spectral fidelity.
	\item[$\bullet$]Analysis of the loss function
	
	The results of different combinations of loss functions are shown in Table \ref{loss}. When $L_{1}$ loss is used, inferior reconstruction performance is gained. With the addition of $L_{spe}$, the PSNR increases by 0.02, and the SAM and ERGAS decrease by 0.08 and 0.03, respectively. The spectral distortion decreases remarkably as the spatial reconstruction successfully improves. When $L_{gra}$ is applied, the PSNR increases by 0.02, and the SAM decreases by 0.005. Attention to adjacent pixel gradients is effective mainly in the spatial dimension \cite{an2023patch}. When 4 loss functions are utilized simultaneously, the PSNR increases by 0.01, and the SAM decreases by 0.003. According to the literature \cite{wang2023asymmetric}, the collaboration of spatial and spectral loss accounts for texture and band correlations, and the addition of $L_{sstv}$ can promote the SR quality further to obtain the best results in the spatial and spectral domains.
\end{itemize}

\begin{table*}[h!]
	\tiny
	\centering
	
	\resizebox{1\linewidth}{!}{
		\begin{tabu}{lcccc|ccccc}
			\hline
			\specialrule{0em}{1pt}{0pt}
			{WPGB}  & {SEGB}  & {SC} &{HS} & {3D}& {PSNR$\uparrow$} &{SSIM$\uparrow$} &{SAM$\downarrow$} &{RMSE$\downarrow$} &{ERGAS$\downarrow$}\\
			\specialrule{0em}{1pt}{0pt}
			\hline

			\checkmark    &&&& &29.0345	&0.8098	&5.4974	&0.0366	&5.8574   \\
			    &\checkmark &&& &29.0625 &0.8102 &5.4861 &0.0365 &5.8365                   \\
			\checkmark &\checkmark &&& &29.0962	&0.8128	&5.4623	&0.0365	&5.8261              \\
			\checkmark &\checkmark &\checkmark  && &29.1282	&0.8151	&5.4212	&0.0362	&5.7973                   \\
			\checkmark    &\checkmark & &\checkmark & &29.0970	&0.8126	&5.4522	&0.0364	&5.8152 \\
			\checkmark    &\checkmark &\checkmark & \checkmark & &29.1440	&0.8155	&5.3973	&0.0362	&5.7855
			\\
			\checkmark &\checkmark &\checkmark &\checkmark &\checkmark &29.1756	&0.8159	&5.3821	&0.0361	&5.7603
			\\
			
			\specialrule{0em}{1pt}{0pt}
			\hline
			
		\end{tabu}
	}
	\caption{Ablation study quantitative comparisons among different components over the testing set of the Pavia dataset at the scale factor of 4.}
	\label{ablation}
\end{table*}

\begin{table*}[h!]
	\tiny
	\centering
	
	\resizebox{0.9\linewidth}{!}{
		\begin{tabu}{lccc|ccccc}
			\hline
			\specialrule{0em}{1pt}{0pt}
			{$L_1$}  & {$L_{spe}$}  & {$L_{gra}$} &{$L_{sstv}$}& {PSNR$\uparrow$} &{SSIM$\uparrow$} &{SAM$\downarrow$} &{RMSE$\downarrow$} &{ERGAS$\downarrow$}\\
			\specialrule{0em}{1pt}{0pt}
			\hline

			\checkmark    &&& &40.3240	&0.9439	&2.3680	&0.0114	&4.9669   \\
			\checkmark &\checkmark && &40.3450	&0.9439	&2.2890	&0.0114	&4.9332                   \\
			\checkmark &\checkmark &\checkmark & &40.3658	&0.9442	&2.2843	&0.0114	&4.9224              \\
			\checkmark &\checkmark &\checkmark  &\checkmark &40.3953 &0.9451 &2.2702 &0.0113	&4.8531                   \\
			
			\specialrule{0em}{1pt}{0pt}
			\hline
			
		\end{tabu}
	}
	\caption{Quantitative performance of different loss functions evaluated over four testing images of Chikusei dataset at the scale factor ×4.}
	\label{loss}
\end{table*}

\section{Conclusion}
In this work, we propose the EFGN to effectively learn the abundant spatial‒spectral consistency for SHSR. First, we use the designed SPDFM to provide multifarious guidance to keep band information consistent by fetching spectral channel and hierarchical spatial information from adjacent subgroups. Benefiting from the gate operations of large kernel convolution and spectral interaction, SSRGM can sufficiently obtain representative spatial–spectral features. Second, we employ 3D-SSRGM to refine the holistic spatial and spectral information. The qualitative and quantitative results on different datasets demonstrated the superiority of our method. The efficiency evaluation demonstrates that EFGN gains high SR quality efficiently, and ablation experiments verify the effectiveness of the modules in the EFGN.
In future, further improvements should be explored. First, the normal group-based methods bring stuffless features as it divides adjacent bands with similar information into subgroup. Second, most convolutions possess fixed kernel size and lack the specific extraction of different objects. And as for the group-based methods, the most process for whole hyperspectral cube are inadequate. In following researches, we would add distant bands with various information into subgroup to increase spectral diversity without breaking the spectral coherence, then the deformable and adaptable convolution could be utilized to obtain diverse feature, and the holistic structure focused on refining overall spatial-spectral information for huge spectral dimension should be designed.





\clearpage 






\bibliographystyle{cas-model2-names}

\bibliography{cas-refs}

\vspace{-5mm}

\bio{wxf}
\textbf{Xufei Wang} received the B. S. degree from QingDao University in 2022, and will receive the M.S. degree in June 2026 from School of Electronic and Information Engineering, Anhui University, majoring in Information and Communication Engineering. He mainly works on low-level computer vision tasks, including image denoising, image super-resolution.
\endbio

\vspace{22mm}

\bio{zmj}
\textbf{Mingjian Zhang} received the B. S. degree from Hunan University of Science and Technology in 2018, and will receive M.S. degree in 2025 from School of Electronic and Information Engineering, Anhui University, majoring in Information and Communication Engineering. He mainly works on image super-resolution.
\endbio

\vspace{22mm}

\bio{gf}
\textbf{Fei Ge} received the B. S. degree from Anhui Agricultural University, will receive the M.S. degree in June 2026 from School of Electronic and Information Engineering, Anhui University, majoring in Electronic Information. He mainly works on low-level computer vision tasks, including image deblurring, image super-resolution.
\endbio

\vspace{40mm}

\bio{zjc}
\textbf{Jinchen Zhu} will receive the M.S. degree in June 2025 from School of Electronic and Information Engineering, Anhui University, majoring in Information and Communication Engineering. He mainly works on low-level computer vision tasks, including image denoising, image super-resolution.
\endbio

\vspace{22mm}

\bio{sw}
\textbf{Wen Sha} is the professor at the School of Artificial Intelligence, Anhui University, and the director of Intelligent Manufacturing Research Center. He is mainly engaged in embedded system development and industrial internet. 
\endbio

\vspace{22mm}

\bio{rjf}
\textbf{JiFeng Ren} will receive the B. S. degree from Anhui University in 2027, majoring in Electronic science and technology. He mainly works on image super-resolution.
\endbio

\vspace{25mm}

\bio{hzm}
\textbf{Zhimeng Hou} is currently pursuing B. S. degree in electronic information engineering at School of Electronic and Information Engineering, Anhui University, Hefei, China. Her research interests include hyperspectral fruit quality detection and image super-resolution. 
\endbio

\vspace{20mm}

\bio{zsg}
\textbf{Shouguo Zheng} received his PhD from University of Chinese Academy of Sciences, is the associate research fellow in Hefei Institutes of Physical Science, Chinese Academy of Sciences. He is mainly engaged in optical detection and image super-resolution. 
\endbio

\vspace{13mm}

\bio{zl}
\textbf{Ling Zheng} is a lecturer at the School of Electronic Information Engineering, Anhui University. She is mainly engaged in machine vision and image processing.  
\endbio

\vspace{28mm}

\bio{weng}
\textbf{Shizhuang Weng} (IEEE Memeber)  received his PhD from University of Science and Technology of China, is an associate professor at the School of Electronic Information Engineering from Anhui University, and a member of the Chinese Society of Artificial Intelligence. He is engaged in computer vision, image processing and deep learning.  
\endbio

\end{document}